\documentclass[12pt]{article}
\usepackage{amsmath, amssymb, amsthm, float}
\usepackage{tikz}
\usepackage{graphicx,psfrag,epsf}
\usepackage{enumerate}
\usepackage{hyperref}
\usepackage{natbib}
\usepackage{url} 

\definecolor{PennRed}{RGB}{152, 30 50}
\definecolor{PennBlue}{RGB}{0, 44, 119}
\definecolor{PennGreen}{RGB}{94, 179,70}
\definecolor{PennViolet}{RGB}{141, 76, 145}
\definecolor{PennSkyBlue}{RGB}{14, 118, 188}
\definecolor{PennOrange}{RGB}{243, 117, 58}
\definecolor{PennBrightRed}{RGB}{223,82, 78}

\hypersetup{pdfborder = {0 0 0.5 [3 3]}, colorlinks = true, linkcolor = PennBrightRed, citecolor = PennSkyBlue}

\usepackage{amssymb, amsthm,multirow,float, subcaption, xcolor, setspace}

\newcommand{\blind}{0}

\newtheorem{theorem}{Theorem}

\newtheorem{proposition}[theorem]{Proposition}

\DeclareMathOperator*{\argmax}{arg\,max}
\DeclareMathOperator*{\argmin}{arg\,min}

\begin{document}

\def\C{\mathbb{C}}
\def\R{\mathbb{R}}
\def\Q{\mathbb{Q}}
\def\Z{\mathbb{Z}}
\def\N{\mathbb{N}}

\def\P{\mathbb{P}}
\def\E{\mathbb{E}}

\def\bgamma{\boldsymbol{\gamma}}
\def\bomega{\boldsymbol{\omega}}
\def\btheta{\boldsymbol{\theta}}
\def\bPhi{\boldsymbol{\Phi}}
\def\bdelta{\boldsymbol{\delta}}
\def\brho{\boldsymbol{\rho}}
\def\bs{\boldsymbol{s}}

\def\by{\mathbf{y}}
\def\bx{\mathbf{x}}
\def\bz{\mathbf{z}}

\def\bY{\mathbf{Y}}
\def\bX{\mathbf{X}}

\def\e{\text{e}}

\def\spacingset#1{\renewcommand{\baselinestretch}%
{#1}\small\normalsize} \spacingset{1}


\if0\blind
{
  \title{\bf Simultaneous Variable and Covariance Selection with the Multivariate Spike-and-Slab LASSO}
  \author{Sameer K. Deshpande\\
    Department of Statistics, University of Pennsylvania\\
    Veronika Ro\v{c}kov\'{a} \\
    Booth School of Business, University of Chicago \\
    and \\
    Edward I. George  \thanks{
    The authors gratefully acknowledge NSF grant DMS-1406563}\hspace{.2cm} \\
    Department of Statistics, University of Pennsylvania}
  \maketitle
} \fi
\if1\blind
{
  \bigskip
  \bigskip
  \bigskip
  \begin{center}
    {\LARGE \bf Simultaneous Variable and Covariance Selection with the Multivariate Spike-and-Slab LASSO}
\end{center}
  \medskip
} \fi

\bigskip
\begin{abstract}
We propose a Bayesian procedure for simultaneous variable and covariance selection using continuous spike-and-slab priors in multivariate linear regression models where $q$ possibly correlated responses are regressed onto $p$ predictors. 
Rather than relying on a stochastic search through the high-dimensional model space, we develop an ECM algorithm similar to the EMVS procedure of Ro\v{c}kov\'{a} \& George (2014) targeting modal estimates of the matrix of regression coefficients and residual precision matrix. 
Varying the scale of the continuous spike densities facilitates dynamic posterior exploration and allows us to filter out negligible regression coefficients and partial covariances gradually. 
Our method is seen to substantially outperform regularization competitors on simulated data. 
We demonstrate our method with a re-examination of data from a recent observational study of the effect of playing high school football on several later-life cognition, psychological, and socio-economic outcomes.

\end{abstract}

\noindent%
\vfill

\newpage
\spacingset{1.45} 
\section{Introduction}
\label{sec:intro}

We consider the multivariate Gaussian linear regression model, in which one simultaneously regresses $q > 1$ possibly correlated responses onto a common set of $p$ covariates.
In this setting, one observes $n$ independent pairs of data $\left(\bx_{i}, \by_{i}\right)$ where $\by_{i} \in \R^{q}$ contains the $q$ outcomes and $\bx_{i} \in \R^{p}$ contains measurements of the covariates.
One then models $\by_{i} = \bx_{i}'B + \varepsilon_{i},$ with $\varepsilon_{1}, \ldots, \varepsilon_{n} \sim \text{N}\left(\mathbf{0}_{q}, \Omega^{-1}\right),$ independently, where $B = \left(\beta_{j,k}\right)_{j,k}$ and $\Omega = \left(\omega_{k,k'}\right)_{k,k'}$ are unknown $p \times q$ and $q \times q$ matrices, respectively. 
The main thrust of this paper is to propose a new methodology for the simultaneous identification of the regression coefficient matrix $B$ and the residual precision matrix $\Omega.$
Our framework additionally includes estimation of $B$ when $\Omega$ is known and estimation of $\Omega$ when $B$ is known as important special cases.

The identification and estimation of a sparse set of regression coefficients has been extensively explored in the univariate linear regression model, often through a penalized likelihood framework.
Perhaps the most prominent method is \citet{Tibshirani1996}'s LASSO, which adds an $\ell_{1}$ penalty to the negative log-likelihood. 
The last two decades have seen a proliferation of alternative penalties, including the adaptive LASSO \citep{Zou2006}, smoothly clipped absolute deviation (SCAD), \citep{FanLi2001}, and minimum concave penalty \citep{Zhang2010}.
Given the abundance of penalized likelihood procedures for univariate regression, when moving to the multivariate setting, it  is very tempting to deploy one's favorite univariate procedure to each of the $q$ responses separately, thereby assembling an estimate of $B$ column-by-column. 
Such an approach fails to account for the correlations between responses and may lead to poor predictive performance (see, e.g., \citet{BreimanFriedman1997}). 
In many applied settings one may reasonably believe that some groups of covariates are simultaneously ``relevant'' to many responses.
A response-by-response approach to variable selection fails to investigate or leverage such structural assumptions.
This has led to the the block-structured regularization approaches of \citet{Turlach2005}, \citet{ObozinskiWainwrightJordan2011} and \citet{PengZhu2010}, among many others.
While these proposals frequently yield highly interpretable and useful models, they do not explicitly model the residual correlation structure, essentially assuming that $\Omega = I.$

Estimation of a sparse precision matrix from multivariate Gaussian data has a similarly rich history, dating back to \citet{Dempster1972}, who coined the phrase \textit{covariance selection} to describe this problem. 
While \citet{Dempster1972} was primarily concerned with estimating the covariance matrix $\Sigma = \Omega^{-1}$ by first sparsely estimating the precision matrix $\Omega,$ recent attention has focused on estimating the underlying Gaussian graphical model, $G$.
The vertices of the graph $G$ correspond to the coordinates of the multivariate Gaussian vector and an edge between vertices $k$ and $k'$ signifies that the corresponding coordinates are conditionally dependent.
These conditional dependency relations are encoded in the support of $\Omega.$ 
A particularly popular approach to estimating $\Omega$ is the graphical LASSO (GLASSO), which adds an $\ell_{1}$ penalty to the negative log-likelihood of $\Omega$ (see, e.g., \citet{YuanLin2007}, \citet{Banerjee2008}, and \citet{Friedman2008}).

While variable selection and covariance selection each have long, rich histories, joint variable and covariance selection has only recently attracted attention. 
To the best of our knowledge, \citet{RothmanLevinaZhu2010} was among the first to consider the simultaneous sparse estimation of $B$ and $\Omega$, solving the penalized likelihood problem:
\begin{equation}
\label{eq:mrce_objective}
\argmin_{B,\Omega}\left\{-\frac{n}{2}\log{\left\lvert \Omega \right\rvert} + \frac{1}{2}\text{tr}\left( \left(\bY - \bX B\right)\Omega \left(\bY - \bX B\right)'\right) + \lambda\sum_{j,k}{\left\lvert \beta_{j,k}\right\rvert} + \xi\sum_{k \neq k'}{\left\lvert \omega_{k,k'}\right\rvert}\right\}
\end{equation}
Their procedure, called MRCE for ``Multivariate Regression with Covariance Estimation'', induces sparsity in $B$ and $\Omega$ with separate $\ell_{1}$ penalties and can be viewed as an elaboration of both the LASSO and GLASSO.
Following \citet{RothmanLevinaZhu2010}, several authors have proposed solving problems similar to that in Equation~\eqref{eq:mrce_objective}: \citet{YinLi2011} considered nearly the same objective but with adaptive LASSO penalties, \citet{LeeLiu2012}  proposed weighting each $\left\lvert \beta_{j,k} \right\rvert $ and $\left\lvert\omega_{k,k'}\right\rvert$ individually, and \citet{AbegazWit2013} replaced the $\ell_{1}$ penalties with SCAD penalties.
Though the ensuing joint optimization problem can be numerically unstable in high-dimensions, all of these authors report relatively good performance in estimating $B$ and $\Omega.$
\citet{Cai2013} takes a somewhat different approach, first estimating $B$ in a column-by-column fashion with a separate Dantzig selector for each response and then estimating $\Omega$ by solving a constrained $\ell_{1}$ optimization problem.
Under mild conditions, they established the asymptotic consistency of their two-step procedure, called CAPME for ``Covariate-Adjusted Precision Matrix Estimation.''

\begin{figure}[H]
\centering
\includegraphics[width = 0.9\textwidth]{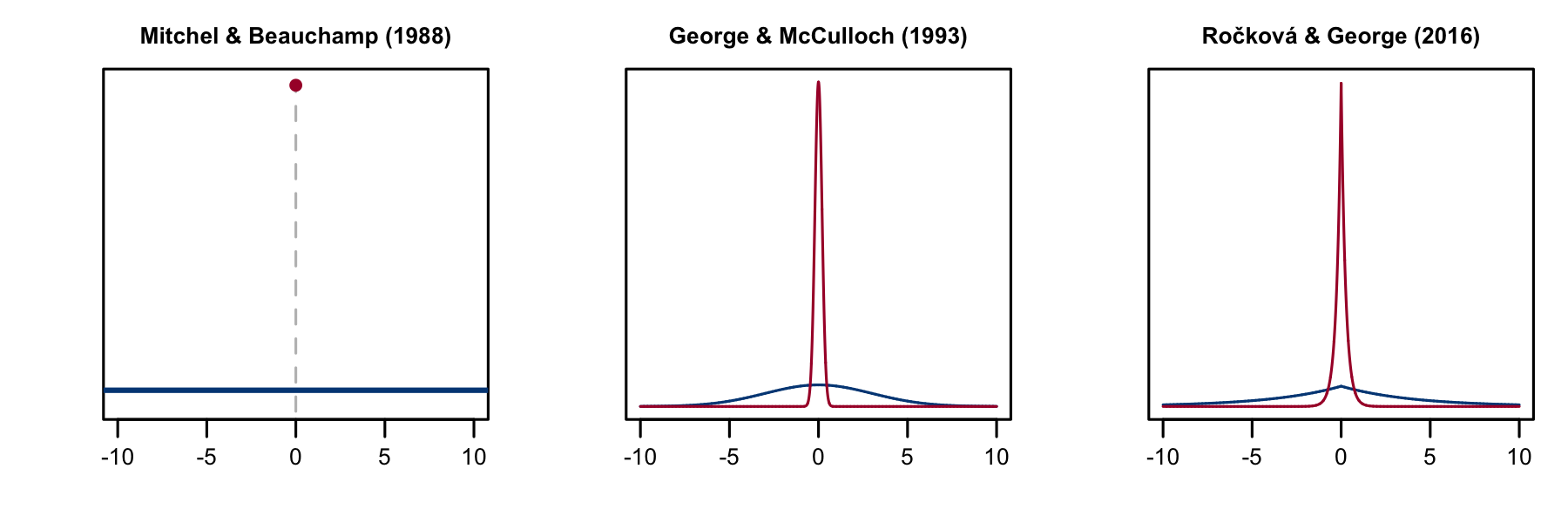}
\caption{Three choices of spike and slab densities. Slab densities are colored red and spike densities are colored blue. The heavier Laplacian tails of \citet{RockovaGeorge2016}'s slab distribution help stabilize non-zero parameter more so than \citet{GeorgeMcCulloch1993}'s Gaussian slabs.}
\label{fig:spike_slab_density}
\end{figure}

Bayesian too have considered variable and covariance selection.
A workhorse of sparse Bayesian modeling is the spike-and-slab prior, in which one models parameters as being drawn \textit{a priori} from either a point-mass at zero (the ``spike'') or a much more diffuse continuous distribution (the ``slab'') \citep{MitchellBeauchamp1988}.
To deploy such a prior, one introduces a latent binary variable for each regression coefficient indicating whether it was drawn from the spike or slab distribution and uses the posterior distribution of these latent parameters to perform variable selection.\cite{GeorgeMcCulloch1993} relaxed this formulation slightly by taking the spike and slab distributions to be zero-mean Gaussians, with the spike distribution very tightly concentrated around zero.
Their relaxation facilitated a straight-forward Gibbs sampler that forms the backbone of their Stochastic Search Variable Selection (SSVS) procedure for univariate linear regression. 
While continuous spike and slab densities generally preclude exactly sparse estimates, the intersection point of the two densities can be viewed as an \textit{a priori} ``threshold of practical relevance.''
More recently, \citet{RockovaGeorge2016} took both the spike and slab distributions to be Laplacian, which led to posterior distributions with exactly sparse modes.
Under mild conditions, their ``spike-and-slab LASSO'' prior produces posterior distributions that concentrate asymptotically around the true regression coefficients at nearly the minimax rate. 
Figure~\ref{fig:spike_slab_density} illustrates these three different spike-and-slab proposals.

An important Bayesian approach to covariance selection begins by specifying a prior over the underlying graph $G$ and a hyper-inverse Wishart prior \citep{DawidLauritzen1993} on $\Sigma | G.$
This prior is constrained to the set of symmetric positive-definite matrices such that off-diagonal entry $\omega_{k,k'}$ of $\Sigma^{-1} = \Omega$ is non-zero if and only if there is an edge between vertices $k$ and $k'$ in $G.$   
See \citet{GiudiciGreen1999}, \citet{Roverato2002}, and \citet{CarvalhoScott2009} for additional methodological and theoretical details on these priors and see \citet{Jones2005} and \citet{CarvalhoMassamWest2007} for computational considerations.
Recently, \citet{HWang2015} and \citet{BanerjeeGhosal2015} placed spike-and-slab priors on the off-diagonal elements of $\Omega,$ using a Laplacian slab and a point-mass spike at zero. 
\citet{BanerjeeGhosal2015} established the posterior consistency in the asymptotic regime where $(q + s)\log{q} = o(n)$ where $s$ is the total number of edges in $G.$

Despite their conceptual elegance, spike-and-slab priors result in highly multimodal posteriors that can slow the mixing of MCMC simulations.
This is exacerbated in the multivariate regression setting, especially when $p$ and $q$ are moderate-to-large relative to $n.$
To overcome this slow mixing when extending SSVS to the multivariate linear regression model, \citet{BrownVannucciFearn1998} restricted attention to models in which a variable was selected as ``relevant'' to either all or none of the responses.
This enabled them to marginalize out the parameter $B$ and directly Gibbs sample the latent spike-and-slab indicators.
Despite the computational tractability, the focus to models in which a covariate affects all or none of the responses may be unrealistic and overly restrictive. 
More recently, \citet{Richardson2010} overcame this by using an evolutionary MCMC simulation, but made the equally restrictive and unrealistic assumption that $\Omega$ was diagonal. 
\citet{BhadraMallick2013} placed spike-and-slab priors on the elements of $B$ and a hyper inverse Wishart prior on $\Sigma | G.$ 
To ensure quick mixing of their MCMC, they made the same restriction as \citet{BrownVannucciFearn1998}: a variable was selected as relevant to all of the $q$ responses or to none of them.
It would seem, then, that a Bayesian who desires a computationally efficient procedure must choose between having a very general sparsity structure in $B$ at the expense of a diagonal $\Omega$ (\`{a} la \citet{Richardson2010}), or a general sparsity structure in $\Omega$ with a peculiar sparsity pattern in $B$ (\`{a} la \citet{BrownVannucciFearn1998} and \citet{BhadraMallick2013}). 
Although their non-Bayesian counter-parts are not nearly as encumbered, the problem of picking appropriate penalty weights via cross-validation can be computationally burdensome.

In this paper, we attempt to close this gap, by extending the EMVS framework of \citet{RockovaGeorge2014} and spike-and-slab LASSO framework of \citet{RockovaGeorge2016} to the multivariate linear regression setting.
EMVS is a deterministic alternative to the SSVS procedure that avoids posterior sampling by targeting local modes of the posterior distribution with an EM algorithm that treats the latent spike-and-slab indicator variables as ``missing data.''
Through its use of Gaussian spike and slab distributions, the EMVS algorithm reduces to solving a sequence of ridge regression problems whose penalties \textit{adapt} to the evolving estimates of the regression parameter.
Subsequent development in \citet{RockovaGeorge2016} led to the spike-and-slab LASSO procedure, in which both the spike and slab distributions were taken to be Laplacian. 
This framework allows us to ``cross-fertilize'' the best of the Bayesian and non-Bayesian approaches: by targeting posterior modes instead of sampling, we may lean on existing highly efficient algorithms for solving penalized likelihood problems while the Bayesian machinery facilities adaptive penalty mixing, essentially for free.

Much like \citet{RockovaGeorge2014}'s EMVS, our proposed procedure reduces to solving a series of penalized likelihood problems.
Our prior model of the uncertainty about which covariate effects and partial residual covariances are large and which are essentially negligible allows us to perform selective shrinkage, leading to vastly superior support recovery and estimation performance compared to non-Bayesian procedures like MRCE and CAPME.
Moreover, we have found our joint treatment of $B$ and $\Omega,$ which embraces the residual correlation structure from the outset, is capable of identifying weaker covariate effects than two-step procedures that first estimate $B$ either column-wise or by assuming $\Omega = I$ and then estimate $\Omega.$

The rest of this paper is organized as follows.
We formally introduce our model and algorithm in Section~\ref{sec:model}.
In Sections~\ref{sec:dpe}, we embed this algorithm within a path-following scheme that facilitates \textit{dynamic posterior exploration}, identifying putative modes of $B$ and $\Omega$ over a range of different posterior distributions indexed by the ``tightness'' of the prior spike distributions.
We present the results of several simulation studies in Section~\ref{sec:simulations}.
In Section~\ref{sec:real_data}, we re-analyze the data of \citet{DeshpandeHasegawa2017}, a recent observational study on the effects of playing high school football on a range of cognitive, behavioral, psychological, and socio-economic outcomes later in life.
We conclude with a discussion in Section~\ref{sec:discussion}.

\section{Model and Algorithm}
\label{sec:model}
We begin with some notation.
We let $\lVert B \rVert_{0}$ be the number of non-zero entries in the matrix $B$ and, abusing the notation somewhat, we let $\lVert \Omega \rVert^{*}_{0}$ be the number of non-zero, off-diagonal entries in the upper triangle of the precision matrix $\Omega.$
For any matrix of covariates effects $B,$ we let $\mathbf{R}(B) = \bY - \bX B$ denote the residual matrix whose $\text{k}^{\text{th}}$ column is denoted $\mathbf{r}_{k}(B).$
Finally, let $S(B) = n^{-1}\mathbf{R}(B)'\mathbf{R}(B)$ be the residual covariance matrix.
In what follows, we will usually suppress the dependence of $\mathbf{R}(B)$ and $S(B)$ on $B,$ writing only $\mathbf{R}$ and $S.$
Additionally, we assume that the columns of $\bX$ have been centered and scaled to have mean 0 and Euclidean norm $\sqrt{n}$ and that the columns of $\bY$ have been centered and are on approximately similar scales.

Recall that our data likelihood is given by
$$
p(\bY | B, \Omega) \propto \left\lvert \Omega\right\rvert^{\frac{n}{2}}\exp{\left\{-\frac{1}{2}\text{tr}\left(\left(\bY - \bX B\right)\Omega\left(\bY - \bX B\right)'\right)\right\}}
$$
We introduce latent 0--1 indicators, $\bgamma = \left(\gamma_{j,k}: 1 \leq j \leq p, 1 \leq k \leq q \right)$ so that, independently for $1 \leq j \leq p, 1 \leq k \leq q,$ we have
$$
\pi(\beta_{j,k} | \gamma_{j,k}) \propto \left(\lambda_{1}\e^{-\lambda_{1}\left\lvert \beta_{j,k}\right\rvert}\right)^{\gamma_{j,k}}\left(\lambda_{0}\e^{-\lambda_{0}\left\lvert \beta_{j,k}\right\rvert}\right)^{1 - \gamma_{j,k}}.
$$

Similarly, we introduce latent 0--1 indicators, $\bdelta = \left(\delta_{k,k'}: 1 \leq k < k' \leq q \right)$ so that, independently for $1\leq k < k' \leq q,$ we have
$$
\pi(\omega_{k,k'} | \delta_{k,k'}) \propto \left(\xi_{1}\e^{-\xi_{1}\left\lvert \omega_{k,k'}\right\rvert}\right)^{\delta_{k,k'}}\left(\xi_{0}\e^{-\xi_{0}\left\lvert \omega_{k,k'}\right\rvert}\right)^{1 - \delta_{k,k'}}
$$
Recall that in the spike-and-slab framework, the spike distribution is viewed as having \textit{a priori} generated all of the negligible parameter values, permitting us to interpret $\gamma_{j,k} = 0$ as an indication that variable $j$ has an essentially null effect on outcome $k$.
Similarly, we may interpret $\delta_{k,k'} = 0$ to mean that the partial covariance between $\mathbf{r}_{k}$ and $\mathbf{r}_{k'}$ is small enough to ignore. 
To model our uncertainty about $\bgamma$ and $\bdelta,$ we use the familiar beta-binomial prior \citep{ScottBerger2010} : 
\begin{align*}
\gamma_{j,k} | \theta &\stackrel{\text{i.i.d}}{\sim} \text{Bernoulli}(\theta) & \theta &\sim \text{Beta}(a_{\theta}, b_{\theta}) \\
\delta_{k,k'} | \eta &\stackrel{\text{i.i.d}}{\sim} \text{Bernoulli}(\eta) & \eta &\sim \text{Beta}(a_{\eta}, b_{\eta})
\end{align*}
where $a_{\theta}, b_{\theta}, a_{\eta},$ and $b_{\eta}$ are fixed positive constants, and $\bgamma$ and $\bdelta$ are \textit{a priori} independent.
We may view $\theta$ and $\eta$ as measuring the proportion of non-zero entries in $B$ and non-zero off-diagonal elements of $\Omega,$ respectively.

Following the example of \citet{HWang2015} and \citet{BanerjeeGhosal2015}, we place independent exponential $\text{Exp}(\xi_{1})$ priors on the diagonal elements of $\Omega.$
This introduces mild regularization to prevent the diagonal elements $\omega_{k,k}$ from becoming massive.
As \citet{LeeLiu2012} note, such $\Omega$ with massive diagonal entries are undesirable as they correspond to regression models with very little residual variation.
At this point, it is worth noting that the requirement $\Omega$ be positive definite introduces dependence between the $\omega_{k,k'}$'s not currently reflected in the above prior. 
In fact, generally speaking, placing independent spike-and-slab priors on the off-diagonal elements and independent exponential priors along the diagonal leads to considerable prior probability being placed outside the cone of symmetric positive semi-definite matrices. 
In light of this, we complete our prior specification by formally truncating to the space of positive definite matrices so that the conditional prior density of $\Omega | \bdelta$ can be written
$$
\pi(\Omega | \eta) \propto \left(\prod_{k = 1}^{q}{\xi_{1}\e^{-\xi_{1}\omega_{k,k}}}\right) \times \left(\prod_{k < k'}{\left\{\delta_{k,k'}\frac{\xi_{1}}{2}\e^{-\xi_{1}\left\lvert \omega_{k,k'}\right\rvert} + (1 - \delta_{k,k'})\frac{\xi_{0}}{2}\e^{-\xi_{0}\left\lvert \omega_{k,k'}\right\rvert}\right\}}\right) \times \mathbb{I}(\Omega \succ 0)
$$
We note in passing that \citet{HWang2015}, \citet{BanerjeeGhosal2015}, and \citet{Gan2018} employ similar truncation. For compactness, we will suppress the restriction $\mathbb{I}(\Omega \succ 0)$ in what follows.

Before proceeding, we take a moment to introduce two functions that will play a critical role in our optimization strategy. 
Given $\lambda_{1}, \lambda_{0}, \xi_{1}$ and $\xi_{0},$ define the functions $p^{\star}, q^{\star}: \R \times [0,1] \rightarrow [0,1]$ by
\begin{align*}
p^{\star}(x,\theta) &= \frac{\theta\lambda_{1}\e^{-\lambda_{1}\left\lvert x \right\rvert}}{\theta\lambda_{1}\e^{-\lambda_{1}\left\lvert x \right\rvert} + (1 - \theta)\lambda_{0}\e^{-\lambda_{0}\left\lvert x \right\rvert}} \\
q^{\star}(x, \eta) &= \frac{\eta \xi_{1}\e^{-\xi_{1}\left\lvert x \right\rvert}}{\eta \xi_{1}\e^{-\xi_{1}\left\lvert x \right\rvert} + (1 - \eta)\xi_{0}\e^{-\xi_{0}\left\lvert x \right\rvert}}.
\end{align*}
Letting $\Xi$ denote the collection $\left\{B, \theta, \Omega, \eta \right\},$ it is straightforward to verify that $p^{\star}(\beta_{j,k} , \theta) = \E\left[\gamma_{j,k} | \bY, \Xi\right]$ and $q^{\star}(\omega_{k,k'}, \eta) = \E\left[\delta_{k,k'} | \bY, \Xi\right],$ the conditional posterior probabilities that $\beta_{j,k}$ and $\omega_{k,k'}$ were drawn from their respective slab distributions.

Integrating out the latent indicators, $\bgamma$ and $\bdelta,$ the log-posterior density of $\Xi$ is, up to an additive constant, given by
\begin{align}
\begin{split}
\label{eq:log_posterior}
\log{\pi(\Xi | \bY)} &= \frac{n}{2}\log{\left\lvert \Omega\right\rvert} - \frac{1}{2}\text{tr}\left( \left( \bY - \bX B\right)'\left(\bY - \bX B\right)\Omega\right) \\
&+ \sum_{j,k}{\log{\left(\theta\lambda_{1}\e^{-\lambda_{1}\left\lvert \beta_{j,k}\right\rvert} + \left(1 - \theta\right)\lambda_{0}\e^{-\lambda_{0}\left\lvert \beta_{j,k}\right\rvert}\right)}} \\
& + \sum_{k,k'}{\log{\left(\eta\xi_{1}\e^{-\xi_{1}\left\lvert \omega_{k,k'}\right\rvert} + \left(1 - \eta\right)\xi_{0}\e^{-\xi_{0}\left\lvert \omega_{k,k'}\right\rvert}\right)}} - \xi_{1}\sum_{k = 1}^{q}{\omega_{k,k}} \\
&+ (a_{\theta} - 1)\log{\theta} + (b_{\theta} - 1)\log{(1 - \theta)} + (a_{\eta} - 1)\log{\eta} + (b_{\eta} - 1)\log{(1 - \eta)}.
\end{split}
\end{align}

Rather than directly sample from this intractable posterior distribution with MCMC, we maximize the posterior density, seeking $\Xi^{*} = \argmax\left\{\log{\pi(\Xi | \bY)}\right\}.$
Performing this joint optimization is quite challenging, especially in light of the non-convexity of the log-posterior density.
To overcome this, we use an Expectation/Conditional Maximization (ECM) algorithm \citep{MengRubin1993} that treats only the partial covariance indicators $\bdelta$ as ``missing data.''
For the E step of this algorithm, we first compute $q^{\star}_{k,k'} := q^{\star}(\omega_{k,k'}^{(t)}, \eta^{(t)}) = \E\left[\delta_{k,k'} | \bY, \Xi^{(t)}\right]$ given a current estimate $\Xi^{(t)}$ and then consider maximizing the surrogate objective function 
\begin{align*}
\E\left[\log{\pi(\Xi, \bdelta | \bY)} | \Xi^{(t)}\right] &= \frac{n}{2}\log{\left\lvert \Omega\right\rvert} - \frac{1}{2}\text{tr}\left( \left( \bY - \bX B\right)'\left(\bY - \bX B\right)\Omega\right)\\
&+ \sum_{j,k}{\log{\left(\theta\lambda_{1}\e^{-\lambda_{1}\left\lvert \beta_{j,k}\right\rvert} + \left(1 - \theta\right)\lambda_{0}\e^{-\lambda_{0}\left\lvert \beta_{j,k}\right\rvert}\right)}} - \sum_{k,k'}{\xi^{\star}_{k,k'}\left\lvert \omega_{k,k'}\right\rvert} - \xi_{1}\sum_{k = 1}^{q}{\omega_{k,k}} \\ 
&+ (a_{\theta} - 1)\log{\theta} + (b_{\theta} - 1)\log{(1 - \theta)} + (a_{\eta} - 1)\log{\eta} + (b_{\eta} - 1)\log{(1 - \eta)}
\end{align*}
where $\xi^{\star}_{k,k'} = \xi_{1}q^{\star}_{k,k'} + \xi_{0}(1 - q^{\star}_{k,k'}).$
We then perform two CM steps, first updating the pair $(B, \theta)$ while holding $(\Omega, \eta) = (\Omega^{(t)}, \eta^{(t)})$ fixed at its previous value and then updating $(\Omega, \eta)$ while fixing $(B, \theta) = (B^{(t+1)}, \theta^{(t+1)})$ at its new value. 
As we will see shortly, augmenting our log-posterior with the indicators $\bdelta$ facilitates simple updates of $\Omega$ by solving a GLASSO problem.
We do not also augment our log-posterior with the indicators $\bgamma$ as the update of $B$ can be carried out with a coordinate ascent strategy despite the non-convex penalty seen in the second line of Equation~\eqref{eq:log_posterior}.

Holding $(\Omega, \eta) = (\Omega^{(t)}, \eta^{(t)})$ fixed, we update $(B, \theta)$ by solving
\begin{equation}
\label{eq:B_theta_update}
(B^{(t+1)}, \theta^{(t+1)}) = \argmax_{B, \theta}\left\{-\frac{1}{2}\text{tr}\left(\left(\bY - \bX B\right)\Omega\left(\bY - \bX B\right)'\right) + \log{\pi(B | \theta)} + \log{\pi(\theta)} \right\}
\end{equation}
where
$$
\pi(B | \theta) = \prod_{j,k}{\left(\theta \lambda_{1}\e^{-\lambda_{1}\left\lvert \beta_{j,k}\right\rvert} + (1 - \theta)\lambda_{0}\e^{-\lambda_{0}\left\lvert \beta_{j,k}\right\rvert}\right)}.
$$
and $\pi(\theta) \propto \theta^{a_{\theta} - 1}(1 - \theta)^{b_{\theta} -1}.$
We do this in a coordinate-wise fashion, sequentially updating $\theta$ with a simple Newton algorithm and updating $B$ by solving the following problem
\begin{equation}
\label{eq:B_update}
\tilde{B} = \argmax_{B}\left\{-\frac{1}{2}\text{tr}\left(\left(\bY - \bX B\right)\Omega\left(\bY - \bX B\right)'\right) + \sum_{j,k}{\text{pen}(\beta_{j,k} | \theta)} \right\}
\end{equation}
where
$$
pen(\beta_{j,k}| \theta) = \log{\left(\frac{\pi\left(\beta_{j,k}| \theta \right)}{\pi(0 |\theta)}\right)} = -\lambda_{1}\left\lvert \beta_{j,k}\right\rvert + \log{\left(\frac{p^{\star}(\beta_{j,k}, \theta)}{p^{\star}(0,\theta)}\right)}.
$$
Using the fact that the columns of $\bX$ have norm $\sqrt{n}$ and Lemma 2.1 of \citet{RockovaGeorge2016}, the Karush-Kuhn-Tucker condition tell us that 
$$
\tilde{\beta}_{j,k} = n^{-1}\left[ \left\lvert z_{j,k} \right\rvert - \lambda^{\star}(\tilde{\beta}_{j,k},  \theta)\right]_{+}\text{sign}(z_{j,k}),
$$
where
\begin{align*}
z_{j,k} &= n\tilde{\beta}_{j,k} + \sum_{k'}{\frac{\omega_{k,k'}}{\omega_{k,k}}\bx_{j}'\mathbf{r}_{k'}(\tilde{B})} \\
\lambda^{\star}_{j,k} := \lambda^{\star}(\tilde{\beta}_{j,k}, \theta) &= \lambda_{1}p^{\star}(\tilde{\beta}_{j,k}, \theta) + \lambda_{0}(1 - p^{\star}(\tilde{\beta}_{j,k}, \theta)).
\end{align*}
The form of $\tilde{\beta}_{j,k}$ above immediately suggests a coordinate ascent strategy with soft-thresholding to compute $\tilde{B}$ that is very similar to the one used to compute LASSO solutions \citep{Friedman2007}. 
As noted by \citet{RockovaGeorge2016}, however, this necessary characterization of $\tilde{B}$ is sufficient only when posterior is unimodal. 
In general, when $p > n$ and when $\lambda_{0}$ and $\lambda_{1}$ are far apart, the posterior tends to be highly multimodal. 
In light of this, cyclically applying the soft-thresholding operator may terminate at a sub-optimal local mode. 

Arguments in \citet{ZhangZhang2012} and \citet{RockovaGeorge2016} lead immediately to the following refined characterization of $\tilde{B},$ which blends hard- and soft-thresholding.
 \begin{proposition}
The entries in the global mode $\tilde{B} = \left(\tilde{\beta}_{j,k}\right)$ in Equation~\eqref{eq:B_update} satisfy
$$
\tilde{\beta}_{j,k} = 
\begin{cases}
n^{-1}\left[\left\lvert z_{j,k} \right\rvert - \lambda^{\star}(\tilde{\beta}_{j,k}, \theta)\right]_{+}\text{sign}\left(z_{j,k}\right)
& \text{when $\left\lvert z_{j,k} \right\rvert > \Delta_{j,k}$} \\
0 & \text{when $\left\lvert z_{j,k}\right\rvert \leq \Delta_{j,k}$}
\end{cases}
$$
where
$$
\Delta_{j,k} = \inf_{t > 0}\left\{\frac{nt}{2} - \frac{pen(\tilde{\beta}_{j,k},\theta)}{\omega_{k,k}t}\right\}
$$
\end{proposition}

As it turns out, the element-wise thresholds $\Delta_{j,k}$ is generally quite hard to compute but can be bounded, as seen in the following analog to Theorem 2.1 of \citet{RockovaGeorge2016}.
\begin{proposition}
Suppose that $\left(\lambda_{1} - \lambda_{0}\right) > 2\sqrt{n\omega_{k,k}}$ and $\left(\lambda^{\star}(0, \theta) - \lambda_{1}\right)^{2} > -2n\omega_{k,k}p^{\star}(0,\theta).$
Then $\Delta^{L}_{j,k} \leq \Delta_{j,k} \leq \Delta^{U}_{j,k}$ where
\begin{align*}
\Delta^{L}_{j,k} &= \sqrt{-2n\omega_{k,k}^{-1}\log{p^{\star}(0,\theta)} - \omega_{k,k}^{-2}d} + \omega_{k,k}^{-1}\lambda_{1} \\
\Delta^{U}_{j,k} &= \sqrt{-2n\omega_{k,k}^{-1}\log{p^{\star}(0,\theta)}} + \omega_{k,k}^{-1}\lambda_{1}
\end{align*}
where $d = - \left( \lambda^{\star}(\delta_{c_{+}},\theta) - \lambda_{1}\right)^{2} - 2n\omega_{k,k}\log{p^{\star}(\delta_{c_{+}},\theta)}$ and $\delta_{c_{+}}$ is the larger root of $\text{pen}''(x | \theta) = \omega_{k,k}.$
\end{proposition}

Together Propositions 1 and 2 suggest a \textit{refined coordinate ascent} strategy for updating our estimate of $B.$
Namely, starting from some initial value $B^{old},$ we can update $\beta_{j,k}$ with the thresholding rule:
$$
\beta_{j,k}^{new} = \frac{1}{n}\left(\left\lvert z_{j,k} \right\rvert - \lambda^{\star}(\beta_{j,k}^{old},\theta)\right)_{+}\text{sign}(z_{j,k})\mathbb{I}\left(\left\lvert z_{j,k} \right\rvert > \Delta^{U}_{j,k}\right).
$$

Before proceeding, we pause for a moment to reflect on the threshold $\lambda^{\star}_{j,k}$ appearing in the KKT condition and Proposition 1, which evolves alongside our estimates of $B$ and $\theta.$
In particular, when our current estimate of $\beta_{j,k}$ is large in magnitude, the conditional posterior probability that it  was drawn from the slab, $p^{\star}_{j,k},$ tends to be close to one so that $\lambda^{\star}_{j,k}$ is close to $\lambda_{1}.$ 
On the other hand, if it is small in magnitude, $\lambda^{\star}_{j,k}$ tends to be close to the much larger $\lambda_{0}.$
In this way, as our EM algorithm proceeds, it performs \textit{selective shrinkage}, aggressively penalizing small values of $\beta_{j,k}$ without overly penalizing larger values. 
It is worth pointing out as well that $\lambda^{\star}_{j,k}$ adapts not only to the current estimate of $B$ but also to the overall level of sparsity in $B$, as reflected in the current estimate of $\theta.$
The adaptation is entirely a product our explicit \textit{a priori} modeling of the latent indicators $\bgamma$ and stands in stark contrast to regularization techniques that deploy fixed penalties. 

Fixing $(\Omega, \eta) = (\Omega^{(t)}, \eta^{(t)})$, we iterate between the refined coordinate ascent for $B$ and the Newton algorithm for $\theta$ until some convergence criterion is reached at some new estimate $(B^{(t+1)}, \theta^{(t+1)}).$
Then, holding $(B, \theta) = (B^{(t+1)}, \theta^{(t+1)}),$ we turn our attention to $\left(\Omega, \eta\right)$ and solving the posterior maximization problem
\begin{align*}
\left(\Omega^{(t+1)}, \eta^{(t+1)}\right) &= \argmax\left\{\frac{n}{2}\log{\left\lvert \Omega\right\rvert} - \frac{1}{2}\text{tr}\left(S\Omega\right) - \sum_{k < k'}{\xi_{k,k'}^{\star}\left\lvert \omega_{k,k'} \right\rvert } - \xi_{1}\sum_{k = 1}^{q}{\omega_{k,k}} \right.\\
~&+ \left. \log{\eta}\times\left(a_{\eta} - 1 + \sum_{k < k'}{q^{\star}_{k,k'}}\right) + \log{(1 - \eta)} \times \left(b_{\eta} - 1 + \sum_{k < k
}{(1 - q^{\star}_{k,k'})}\right) \right\}.
\end{align*}

It is immediately clear that there is a closed form update of $\eta$:
$$
\eta^{(t+1)} = \frac{a_{\eta} - 1 + \sum_{k < k'}{q^{\star}_{k,k'}}}{a_{\eta} + b_{\eta} - 2 + q(q-1)/2}.
$$
For $\Omega,$ we recognize the M Step update of $\Omega$ as a GLASSO problem.
\begin{equation}
\label{eq:Omega_M_step}
\Omega^{(t+1)} = \argmax_{\Omega \succ 0}\left\{\frac{n}{2}\log{\left\lvert \Omega\right\rvert} - \frac{n}{2}\text{tr}\left( S\Omega\right) - \sum_{k < k'}{\xi_{k,k'}^{\star}\left\lvert \omega_{k,k'}\right\rvert} - \xi_{1}\sum_{k = 1}^{q}{\omega_{k,k}}\right\}
\end{equation}
To find $\Omega^{(t+1)},$ rather than using the block-coordinate ascent algorithms of \citet{Friedman2008} and \citet{Witten2011}, we use the state-of-art QUIC algorithm of \citet{Hsieh2014}, which is based on a quadratic approximation of the objective function and achieves a super-linear convergence rate.
Each of these algorithms returns a positive semi-definite $\Omega^{(t+1)}.$
Just like with the $\lambda^{\star}_{j,k}$'s, the penalties $\xi^{\star}_{k,k'}$ in Equation~\eqref{eq:Omega_M_step} adapt to the values of the current estimates of $\omega_{k,k'}$ and the overall level of sparsity in $\Omega,$ captured by $\eta.$ 

In our implementation, we iterate between the E and CM steps until the percentage change in every $\beta_{j,k}$ and $\omega_{k,k'}$ estimate is less than a user-specified threshold (e.g. $10^{-3}$ or $10^{-6}$) or if the percentage increase in objective value is less than that same threshold.
Because of the non-convexity of our log-posterior, there are no theoretical guarantees that our algorithm terminates at a global mode. 
Indeed, with our stopping criterion, the most we can say is that it will terminate in the vicinity of a stationary point.

Finally, we note that our proposed framework for simultaneous variable and covariance selection can easily be modified to estimate $B$ when $\Omega$ is known and to estimate $\Omega$ when $B$ is known. 
Concurrently with but independently of us, \citet{LiMcCormick2017} and \citet{Gan2018} have considered graphical model estimation with spike-and-slab priors. Specifically, \citet{LiMcCormick2017} use Gaussian spike and slabs that facilitate closed form updates in the the CM steps of their ECM algorithm. Like us, \citet{Gan2018} have used spike-and-slab LASSO priors in an EM algorithm. They have also provided theoretical results showing that $\ell_{\infty}$ approximation error of $\Omega$ is $O\left(\sqrt{\frac{\log{q}}{n}}\right)$ under certain conditions.

\section{Dynamic Posterior Exploration}
\label{sec:dpe}

Given any specification of hyper-parameters $\left(a_{\theta}, b_{\theta}, a_{\eta}, b_{\eta}\right)$ and $\left(\lambda_{1}, \lambda_{0}, \xi_{1}, \xi_{0}\right),$ it is straightforward to deploy the ECM algorithm described in the previous section to identify a putative posterior mode.
We may moreover run our algorithm over a range of hyper-parameter settings to estimate the mode of a range of different posteriors.
Unlike MCMC, which expends considerable computational effort sampling from a single posterior, this \textit{dynamic posterior exploration} provides a snapshot of several different posteriors.

\begin{figure}[H]
\centering
\includegraphics[width = 0.6\textwidth]{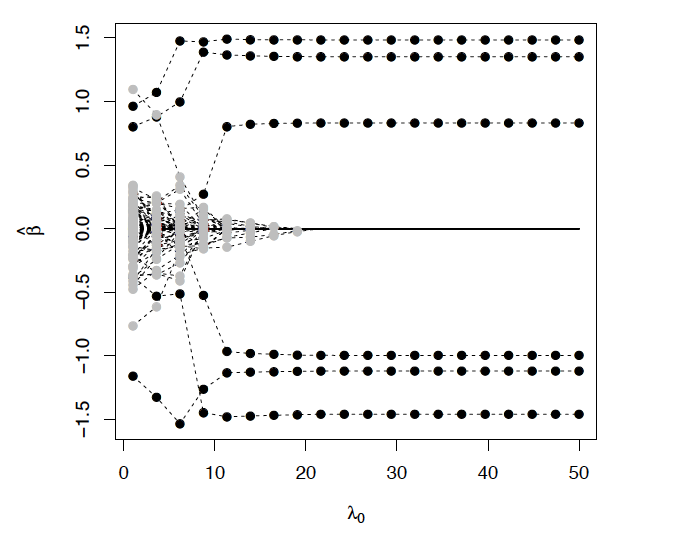}
\caption{Trajectory of parameter estimates in \citet{RockovaGeorge2016}'s dynamic posterior exploration.}
\label{fig:ssl_trajectory}
\end{figure}

In the univariate regression setting, \citet{RockovaGeorge2016} proposed a path-following scheme in which they fixed $\lambda_{1}$ and identified modes of a range of posteriors indexed by a ladder of increasing $\lambda_{0}$ values, $\mathcal{I}_{\lambda} =  \{\lambda_{0}^{(1)} <  \cdots < \lambda_{0}^{(L)}\}$ with sequential re-initialization to produce a sequence of posterior modes.
To find the mode corresponding to $\lambda_{0} = \lambda^{(s)}_{0},$ they ``warm started'' from the previously discovered mode corresponding to $\lambda_{0} = \lambda^{(s-1)}_{0}.$
Early in this path-following scheme, when $\lambda_{0}$ is close to $\lambda_{1},$ distinguishing relevant parameters from negligible is difficult as the spike and slab distributions are so similar.
As $\lambda_{0}$ increases, however, the spike distribution increasingly absorbs the negligible values and results in sparser posterior modes.
Remarkably, \citet{RockovaGeorge2016} found that the trajectories of individual parameter estimates tended to stabilize relatively early in the path, indicating that the parameters had cleanly segregated into groups of zero and non-zero values.
This is quite evident in Figure~\ref{fig:ssl_trajectory} (a reproduction of Figure 2c of \citet{RockovaGeorge2016}), which shows the trajectories of several parameter estimates as a function of $\lambda_{0}.$ 

The stabilization evident in Figure~\ref{fig:ssl_trajectory} allowed them to focus on and report a single model out of the $L$ that they computed without the need for cross-validation. 
From a practitioner's point of view, the stabilization of the path-following scheme sidesteps the issue of picking just the right $\lambda_{0}$: one may specify a ladder spanning a wide range of $\lambda_{0}$ values and observe whether or not the trajectories stabilize after a certain point. 
If so, one may then report any stable estimate and if not, one can expand the ladder to include even larger values of $\lambda_{0}.$
It may be helpful to compare dynamic posterior exploration pre-stabilization to focusing a camera lens: starting from a blurry image, turning the focus ring slowly brings an image into relief, with the salient features becoming increasingly prominent.
In this way, the priors serve more as filters for the data likelihood than as encapsulations of any real subjective beliefs. 

Building on this dynamic posterior exploration strategy for our multivariate setting, we begin by specifying ladders $\mathcal{I}_{\lambda} = \{\lambda_{0}^{(1)} < \cdots < \lambda_{0}^{(L)}\}$ and $\mathcal{I}_{\xi} = \{\xi_{0}^{(1)} < \cdots < \xi_{0}^{(L)}\}$ of increasing $\lambda_{0}$ and $\xi_{0}$ values.
We then identify a sequence $\{\hat{\Xi}^{s,t}: 1 \leq s, t \leq L \},$ where $\hat{\Xi}^{s,t}$ is an estimate of the mode of the posterior corresponding to the choice $(\lambda_{0}, \xi_{0}) = (\lambda_{0}^{(s)}, \xi_{0}^{(t)}),$ which we denote $\Xi^{s,t*}.$
When it comes time to estimate $\Xi^{s,t*},$ we launch our ECM algorithm from whichever of $\hat{\Xi}^{s-1,t}, \hat{\Xi}^{s,t-1}$ and $\hat{\Xi}^{s-1,t-1}$ has the largest log-posterior density, computed according to Equation~\eqref{eq:log_posterior} with $\lambda_{0} = \lambda_{0}^{(s)}$ and $\xi_{0} = \xi_{0}^{(t)}.$
We implement this dynamic posterior exploration by starting with $B = \mathbf{0}, \Omega = I$ and looping over the $\lambda_{0}^{s}$ values and $\xi_{0}^{t}$ values. 
Proceeding in this way, we propagate a single estimate of $\Xi$ through a series of prior filters indexed by the pair $(\lambda_{0}^{(s)}, \xi_{0}^{(t)}).$

When $\lambda_{0}$ is close to $\lambda_{1},$ our refined coordinate ascent can sometimes promote the inclusion of many negligible but non-null $\beta_{j,k}$'s.
Such a specification combined with a $\xi_{0}$ that is much larger than $\xi_{1},$ could over-explain the variation in $\bY$ using several covariates, leaving very little to the residual conditional dependency structure and a severely ill-conditioned residual covariance matrix $S.$
In our implementation, we do not propagate any $\hat{\Xi}^{s,t}$ where the accompanying $S$ has condition number exceeding $10n.$
While this choice is decidedly arbitrary, we have found it to work rather well in simulation studies. 
When it comes time to estimate $\Xi^{s,t*},$ if each of $\hat{\Xi}^{s-1,t}, \hat{\Xi}^{s,t-1}$ and $\hat{\Xi}^{s-1,t-1}$ is numerically unstable, we re-launch our EM algorithm from $B = \mathbf{0}$ and $\Omega = I.$

To illustrate this procedure, which we call mSSL-DPE for ``Multivariate Spike-and-Slab LASSO with Dynamic Posterior Exploration,'' we simulate data from the following model with $n = 400, p = 500,$ and $q = 25.$
We generate the matrix $\bX$ according to a $\text{N}_{p}\left(\mathbf{0}_{p}, \Sigma_{X}\right)$ distribution where $\Sigma_{X} = \left(0.7^{\left\lvert j - j' \right\rvert}\right)_{j,j' = 1}^{p}.$
We construct matrix $B_{0}$ with $pq/5$ randomly placed non-zero entires independently drawn uniformly from the interval $[-2,2].$
This allows us to gauge mSSL-DPE's ability to recover signals of varying strength.
We then set $\Omega_{0}^{-1} = \left(0.9^{\left\lvert k - k' \right\rvert}\right)_{k,k' = 1}^{q}$ so that $\Omega_{0}$ is tri-diagonal, with all $\left\lVert \Omega_{0} \right\rVert_{0} = q-1$ non-zero entries immediately above the diagonal.
Finally, we generate data $\bY = \bX B_{0} + E$ where the rows of $E$ are independently $\text{N}\left(\mathbf{0}_{q}, \Omega_{0}^{-1}\right)$.
For this simulation, we set  $\lambda_{0} = 1, \xi_{0} = 0.01n$ and set $I_{\lambda}$  and $I_{\xi}$ to contain $L = 50$ equally spaced values ranging from 1 to $n$ and from $0.1n$ to $n$, respectively.

In order to establish posterior consistency in the univariate linear regression, \citet{RockovaGeorge2016} required the prior on $\theta$ to place most of its probability in a small interval near zero and recommended taking $a_{\theta} = 1$ and $b_{\theta} = p.$
This concentrates their prior on models that are relatively sparse. 
With $pq$ coefficients in $B$, we take $a_{\theta} = 1$ and $b_{\theta} = pq$ for this demonstration. 
We further take $a_{\eta} = 1$ and $b_{\eta} = q,$ so that the prior on the underlying residual Gaussian graph $G$ concentrates on very sparse graphs with average degree just less than one. 
We will examine the sensitivity of our results to these choices in Appendix~\ref{sec:hyper_parameters}.

Figure~\ref{fig:support_trajectory} shows the trajectory of the number of non-zero $\beta_{j,k}$'s and $\omega_{k,k'}$'s identified at a subset of putative modes $\hat{\Xi}^{s,t}.$
Points corresponding to numerically unstable modes were colored red and points corresponding to those $\hat{\Xi}^{s,t}$ for which the estimated supports of $B$ and $\Omega$ were identical to the estimated supports at $\hat{\Xi}^{L,L},$ were colored blue.
Figure~\ref{fig:support_trajectory} immediately suggests a certain stabilization of our multivariate dynamic posterior exploration. 
In addition to looking at $\lVert \hat{B} \rVert_{0}$ and $\lVert \hat{\Omega} \rVert^{*}_{0},$ we can look at the log-posterior density of each $\hat{\Xi}^{s,t}$ computed with $\lambda_{0} = \lambda_{0}^{(L)}, \xi_{0} = \xi_{0}^{(L)}.$
Figure~\ref{fig:log_post_trajectory} plots a heat map of the ratio $\frac{\log{\pi(\hat{\Xi}^{s,t} | \bY)/\pi(\hat{\Xi}^{0,0} | \bY)}}{\log{\pi(\hat{\Xi}^{L,L} | \bY)/\pi(\hat{\Xi}^{0,0} | \bY)}}$. 
It is interesting to note that this ratio appears to stabilize before the supports did. 

\begin{figure}[H]
\begin{subfigure}[b]{0.48\textwidth}
\centering
\includegraphics[width = \textwidth]{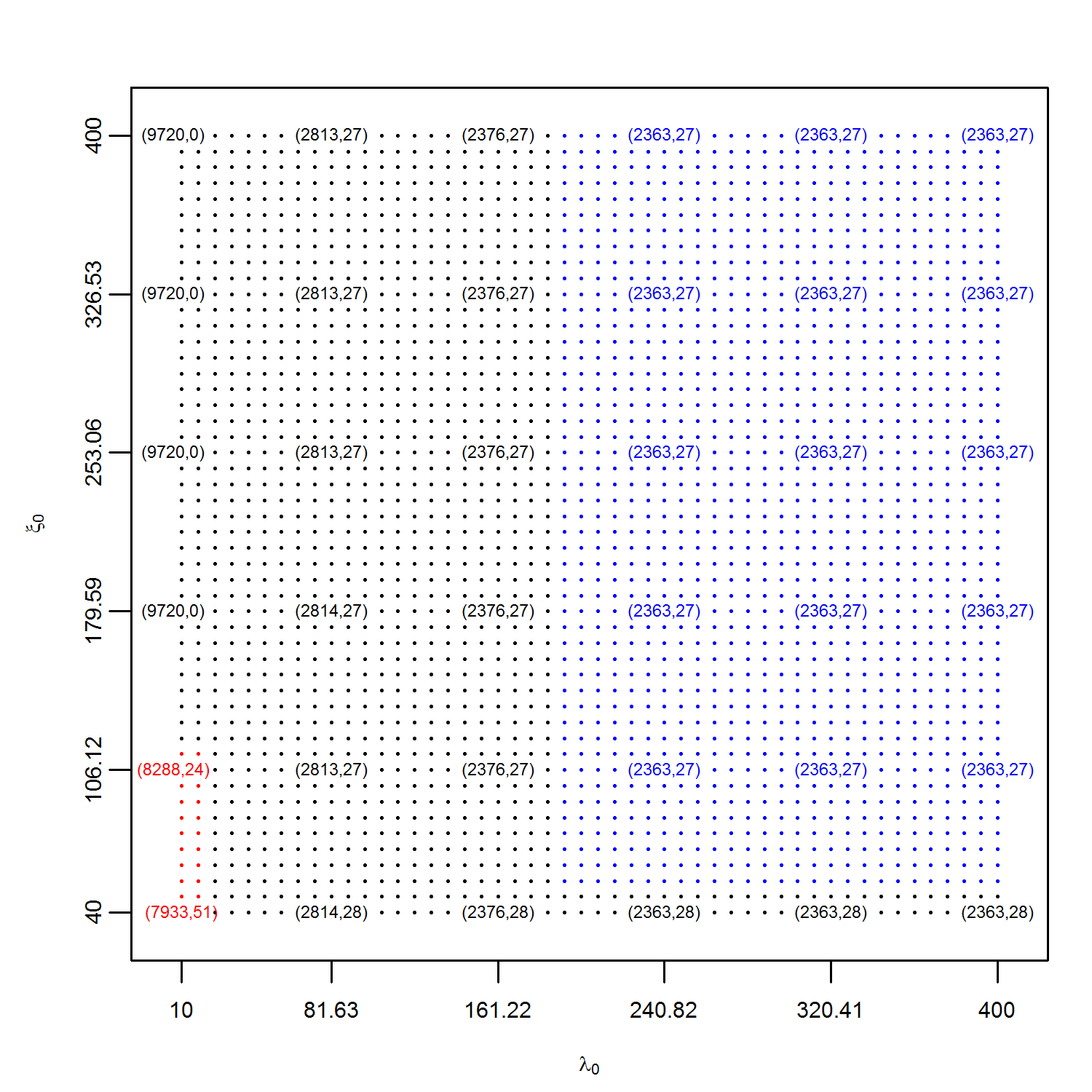}
\caption{}
\label{fig:support_trajectory}
\end{subfigure}
\begin{subfigure}[b]{0.48\textwidth}
\centering
\includegraphics[width = \textwidth]{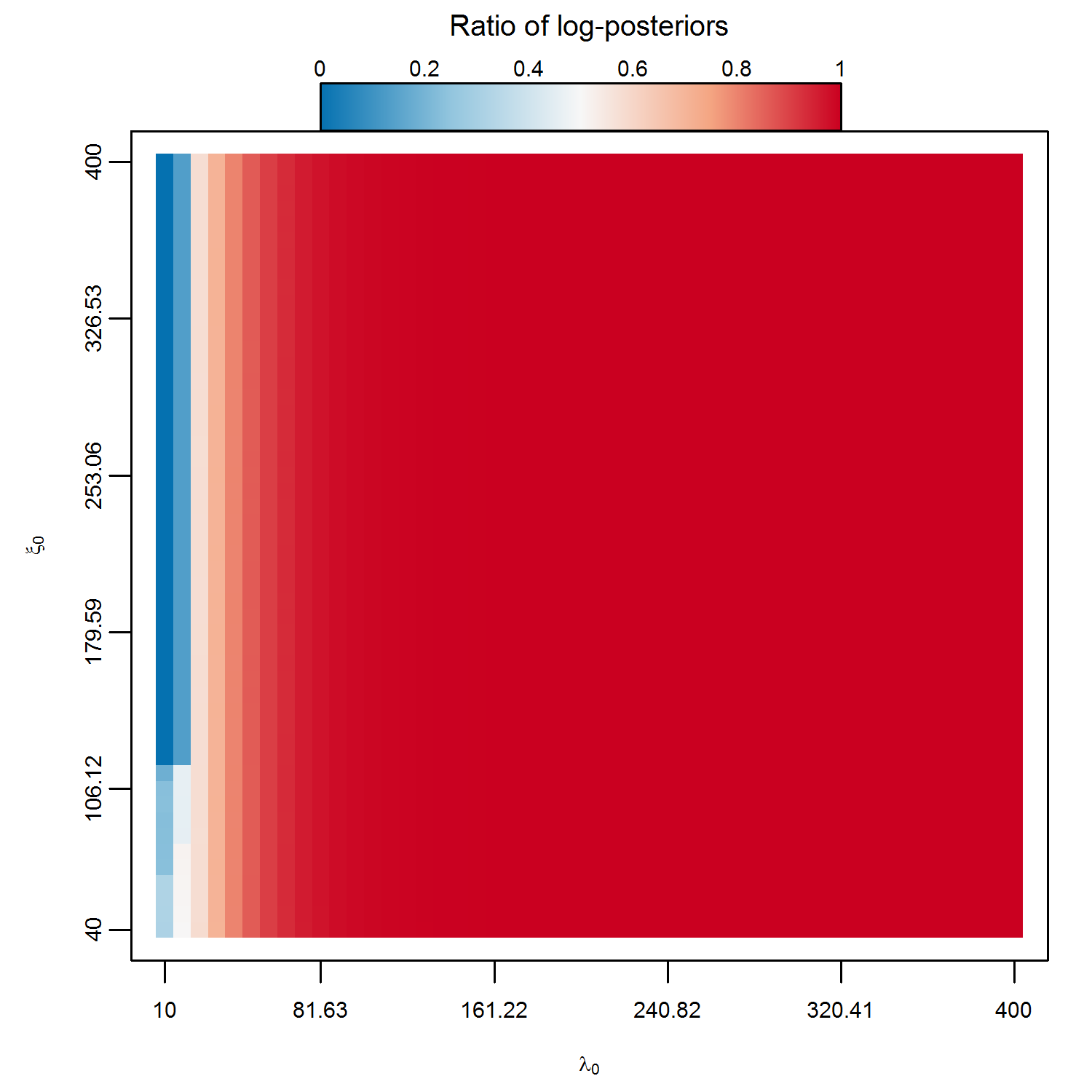}
\caption{}
\label{fig:log_post_trajectory}
\end{subfigure}
\caption{(a)Trajectory of $\left(\lVert B \rVert_{0}, \lVert \Omega \rVert^{*}_{0}\right)$, (b) Trajectory of $\frac{\log{\left(\pi(\hat{\Xi}^{s,t} | \bY)/\pi(\hat{\Xi}^{0,0} | \bY)\right)}}{\log{\left(\pi(\hat{\Xi}^{L,L} | \bY)/\pi(\hat{\Xi}^{0,0} | \bY)\right)}}$}
\label{fig:dpe_trajectories}
\end{figure}

The apparent stabilization in Figure~\ref{fig:dpe_trajectories} allows us to focus on and report a single estimate $\hat{\Xi}^{L,L},$ corresponding to the top-right point in Figure~\ref{fig:support_trajectory}, avoiding costly cross-validation.
Of course, this estimate is nearly indistinguishable from the estimates corresponding to the other blue points in Figure~\ref{fig:support_trajectory} and we could just as easily report any one of them.
On this dataset, mSSL-DPE correctly identified 2360 out of the 2500 non-zero $\beta_{j,k}$'s with only 3 false positives and correctly identified all 24 non-zero $\omega_{k,k'}$'s in the upper triangle of $\Omega,$ again with only 3 false positive identifications. 
We should point out that there is no general guarantee of stabilization for arbitrary ladders $\mathcal{I}_{\lambda}$ and $\mathcal{I}_{\xi}.$
However, in all of the examples we have tried, we found that stabilization occurred long before $\lambda_{0}^{(s)}$ and $\xi_{0}^{(t)}$ reached $\lambda_{0}^{(L)} = \xi_{0}^{(L)} = n.$  

To get a sense as to why such stabilization can occur, suppose that $\lambda_{0}$ is large and that our estimate of $\beta_{j,k}$ is large enough in absolute value that the corresponding $p_{j,k}^{\star}$ is closer to one.
In this case, when we increase $\lambda_{0},$ the new value of $p_{j,k}^{\star}$ in the next iteration of our dynamic posterior exploration will be increasingly closer to one.
This in turn means that the new $\lambda_{j,k}^{\star}$ will decrease towards $\lambda_{0}.$
The ensuing estimate of $\beta_{j,k}$ will increase slightly in magnitude, as a result of $z_{j,k}$ being subjected to a smaller threshold.
On the other hand, if our estimate of $\beta_{j,k}$ is zero, then increasing $\lambda_{0}$ only pushes $p_{j,k}^{\star}$ closer to zero and increases the threshold $\lambda_{j,k}^{\star}$ closer to the new, larger value of $\lambda_{0}.$
Since the threshold increases, the resulting estimate of $\beta_{j,k}$ will remain zero. 
In other words, when $\lambda_{0}$ is large, non-zero estimate of $\beta_{j,k}$ tend to remain non-zero and zero estimates tend to remain zero as we continue to increase $\lambda_{0}.$
Ultimately, the changein the non-zero $\beta_{j,k}$ estimates is dictated by how quickly the corresponding $p_{j,k}^{\star}$ approach one. 
We should add that once the solutions stabilize, the algorithm runs quite quickly so the excess computations are not at all burdensome.

\subsection{Faster Dynamic Conditional Posterior Mode Exploration}
\label{sec:dcpe}

mSSL-DPE can expend considerable computational effort identifying modal estimates $\hat{\Xi}^{s,t}$ corresponding to smaller values of $\lambda_{0}$ and $\xi_{0}.$
Although the support recovery performance of $\hat{\Xi}^{L,L}$ from mSSL-DPE is very promising, one might also consider streamlining the procedure using the following procedure we term mSSL-DCPE for ``Dynamic \textit{Conditional} Posterior Exploration.''
First, we fix $\Omega = I$ and sequentially solve Equation~\eqref{eq:B_theta_update} for each $\lambda_{0} \in I_{\lambda},$ with warm-starts. 
This produces a sequence $\{(\hat{B}^{s}, \hat{\theta}^{s})\}$ of \textit{conditional} posterior modes of $\left(B, \theta\right) | \bY, \Omega = I.$
Then, holding $(B, \theta) = (\hat{B}^{L}_{0}, \hat{\theta}^{L}_{0})$ fixed, we run a modified version of our dynamic posterior exploration to produce a sequence $\{(\hat{\Omega}^{t}, \hat{\eta}^{t})\}$ of conditional modes of $(\Omega, \eta) | \bY, B = \hat{B}^{L}.$ 
We finally run our ECM algorithm starting from $(\hat{B}^{L}, \hat{\theta}^{L}, \hat{\Omega}^{L}, \hat{\eta}^{L})$ with $\lambda_{0} = \lambda_{0}^{L}$ and $\xi_{0}  = \xi_{0}^{L}$ to arrive at an estimate of $\Xi^{L,L*},$ which we denote $\tilde{\Xi}^{L,L}.$
In general, $\tilde{\Xi}^{L,L}$ will be different than the solution obtained by mSSL-DPE, $\hat{\Xi}^{L,L},$ since the two algorithms typically laugh the ECM algorithm from different points when it comes time to estimate $\Xi^{L,L*}.$

In sharp contrast mSSL-DPE, which visits several joint posterior modes before reaching an estimate of posterior mode $\Xi^{L,L*},$ mSSL-DCPE visits several conditional posterior modes to reach another estimate of the same mode.
On the same dataset from the previous subsection, mSSL-DCPE correctly identified 2,169 of the 2,500 non-zero $\beta_{j,k}$ with 8 false positives and all 24 non-zero $\omega_{k,k'}$'s but with 28 false positives.
This was all accomplished in just under 30 seconds, a considerable improvement over the two hour runtime of mSSL-DPE on the same dataset. 
Despite the obvious improvement in runtime, mSSL-DCPE terminated at a sub-optimal point whose log-posterior density was much smaller than the solution found by mSSL-DPE. 
All of the false negative identifications in the support of $B$ made by both procedures corresponded to $\beta_{j,k}$ values which were relatively small in magnitude.
Interestingly, mSSL-DPE was better able to detect smaller signals than mSSL-DCPE. 
We will return to this point later in Section~\ref{sec:simulations}. 

\subsection{Simulations}
\label{sec:simulations}

We now assess the performance of mSSL-DPE and mSSL-DCPE on data simulated from two models, one low-dimensional with $n = 100, p = 50$ and $q =25$ and the other somewhat high-dimensional with $n = 400, p = 500, q =25.$
Just as above, we generate the matrix $\bX$ according to a $\text{N}_{p}\left(\mathbf{0}_{p}, \Sigma_{X}\right)$ distribution where $\Sigma_{X} = \left(0.7^{\left\lvert j - j' \right\rvert}\right)_{j,j' = 1}^{p}.$
We construct matrix $B_{0}$ with $pq/5$ randomly placed non-zero entires independently drawn uniformly from the interval $[-2,2].$
We then set $\Omega_{0}^{-1} = \left(\rho^{\left\lvert k - k' \right\rvert}\right)_{k,k' = 1}^{q}$ for $\rho \in \left\{0, 0.5, 0.7, 0.9\right\}.$
When $\rho \neq 0,$ the resulting $\Omega_{0}$ is tri-diagonal.
Finally, we generate data $\bY = \bX B_{0} + E$ where the rows of $E$ are independently $\text{N}\left(\mathbf{0}_{q}, \Omega_{0}^{-1}\right)$.
For this simulation, we set  $\lambda_{1} = 1, \xi_{1} = 0.01n$ and set $\mathcal{I}_{\lambda}$  and $\mathcal{I}_{\xi}$ to contain $L = 10$ equally spaced values ranging from 1 to $n$ and from $0.1n$ to $n$, respectively.
Like in the previous subsection, we took $a_{\theta} = 1, b_{\theta} = pq, a_{\eta} = 1$ and $b_{\eta} = q.$ We will examine the sensitivity of our results to these hyper-parameters more carefully in Appendix~\ref{sec:hyper_parameters}.

We simulated 100 datasets according to each model, each time keeping $B_{0}$ and $\Omega_{0}$ fixed but drawing a new matrix of errors $E$.
To assess the support recovery and estimation performance, we tracked the following quantities: SEN (sensitivity), SPE (specificity), PREC (precision), ACC (accuracy), MCC (Matthew's Correlation Coefficient), MSE (mean square error in estimating $B_{0}$), FROB (squared Frobenius error in estimating $\Omega_{0}$), and TIME (execution time in seconds).
If we let TP, TN, FP, and FN denote the total number of true positive, true negative, false positive, and false negative identifications made in the support recovery, these quantities are defined as:
\begin{align*}
\text{SEN} &= \frac{\text{TP}}{\text{TP} + \text{FN}} & \text{PREC} &= \frac{\text{TP}}{\text{TP} + \text{FP}} \\
\text{SPE} &= \frac{\text{TN}}{\text{TN} + \text{FP}} & \text{ACC} &= \frac{\text{TP} + \text{TN}}{\text{TP} + \text{TN} + \text{FP} + \text{FN}}
\end{align*}
and
$$
\text{MCC} = \frac{\text{TP} \times \text{TN} - \text{FP} \times \text{FN}}{\sqrt{\left(\text{TP} + \text{FP}\right)\left(\text{TP} + \text{FN}\right)\left(\text{TN} + \text{FP}\right)\left(\text{TN} + \text{FN}\right)}}.
$$

Tables~\ref{tab:low_dim_var} -- ~\ref{tab:hi_dim_cov} report the average performance of mSSL-DPE, mSSL-DCPE, \citet{RothmanLevinaZhu2010}'s MRCE procedure, \citet{Cai2013}'s CAPME procedure, each with 5-fold cross-validation, and the following two competitors:
\begin{description}
\item[Sep.L+G: ]{We first estimate $B$ by solving separate LASSO problems with 10-fold cross-validation for each outcome. We then estimate $\Omega$ from the resulting residual matrix using the GLASSO procedure of \citet{Friedman2008}, also run with 10-fold cross-validation} 
\item[Sep.SSL + SSG:]{We first estimate $B$ column-by-column, deploying \citet{RockovaGeorge2016}'s path-following SSL along the ladder $\mathcal{I}_{\lambda}$ separately for each outcome. We then run a modified version of our dynamic posterior exploration that holds $B$ fixed and only updates $\Omega$ and $\eta$ with the ECM algorithm along the ladder $\mathcal{I}_{\xi}.$ This is similar to \textbf{Sep.L+G} but with adaptive spike-and-slab lasso penalties rather than fixed $\ell_{1}$ penalties.} 
\end{description}

The procedures mSSL-DPE, mSSL-DCPE, and Sep.SSL+SSG are available in the ``SSLASSO'' \texttt{R} package. The simulations were carried out on a high-performance computing cluster, with each node running an Intel Xeon E5-2667 3.30 GHz processor. Each simulated dataset was analyzed on a single core with 5 GB of RAM.


\begin{table}[H]
\centering
\small
\singlespacing
\caption{Average variable selection performance in the low-dimensional setting. MSE has been re-scaled by a factor of 1000. NaN indicates that the specified quantity was undefined, either because no non-zero estimates were returned or because there were truly no non-zero parameters (Simulation 4).}
\label{tab:low_dim_var}
\begin{tabular}{lccccc}\hline
Method & SEN/SPE & PREC/ACC & MCC & MSE & TIME \\ \hline
\multicolumn{6}{c}{Simulation 1: $n = 100, p = 50, q = 25, \rho = 0.9$} \\ \hline
mSSL-DPE &  0.86  / \textbf{1.00} & \textbf{1.00} / 0.97 & \textbf{0.91} & \textbf{1.66} & 10.23 \\
mSSL-DCPE & 0.74 / \textbf{1.00} & 0.99 / 0.95 & 0.82 & 6.69 & 0.37 \\
MRCE & 0.87 / 0.70 & 0.43 / 0.74 & 0.47 & 32.64 & 1467.64 \\
CAPME & \textbf{0.96} / 0.23 & 0.24 / 0.38 & 0.20 & 26.46 & 133.26 \\
SEP.L+G & 0.85 / 0.84 & 0.57 / 0.84 & 0.60& 17.27 & 2.62 \\
SEP.SSL+SSG & 0.73 / \textbf{1.00} & 0.98 / 0.94 & 0.82 & 8.90 & \textbf{0.09} \\ \hline
\multicolumn{6}{c}{Simulation 2: $n = 100, p = 50, q = 25, \rho = 0.7$} \\ \hline
mSSL-DPE & 0.80 / \textbf{1.00} & \textbf{0.99} / \textbf{0.96} & \textbf{0.87} & \textbf{3.53} & 1.62 \\
mSSL-DCPE & 0.72 / \textbf{1.00} & \textbf{0.99} / 0.94 & 0.82 & 7.62 & 0.21 \\
MRCE & \textbf{0.90} / 0.65 & 0.40 / 0.70 & 0.45 & 14.04 & 1704.69 \\
CAPME & 0.86 / 0.74 & 0.47 / 0.77 & 0.50 & 23.89& 137.88 \\
SEP.L+G & 0.85 / 0.84 & 0.56 / 0.84 & 0.60 & 17.43 & 2.60 \\
SEP.SSL+SSG & 0.73 / \textbf{1.00} & \textbf{0.99} / 0.94 & 0.82 & 8.67 & \textbf{0.07} \\ \hline
\multicolumn{6}{c}{Simulation 3: $n = 100, p = 50, q = 25, \rho = 0.5$} \\ \hline
mSSL-DPE & 0.76 / \textbf{1.00} & \textbf{0.99} / \textbf{0.95} & \textbf{0.84} & \textbf{6.02} & 1.28 \\
mSSL-DCPE & 0.73 / \textbf{1.00} & \textbf{0.99} /  0.94 & 0.82 & 8.68 & 0.16 \\
MRCE &  \textbf{0.91} / 0.65 & 0.40 / 0.71 & 0.45 & 10.04 & 714.19 \\
CAPME & 0.86 / 0.76 & 0.47 / 0.78 & 0.52 & 23.54 & 138.30 \\
SEP.L+G & 0.85 / 0.84 & 0.57 / 0.84 & 0.60 & 17.32 & 2.56 \\
SEP.SSL+SSG & 0.73 / \textbf{1.00} & \textbf{0.99} / 0.94 & 0.82 & 8.59 & \textbf{0.06} \\ \hline
\multicolumn{6}{c}{Simulation 4: $n = 100, p = 50, q = 25, \rho = 0$} \\ \hline
mSSL-DPE & 0.73 / \textbf{1.00} & \textbf{0.99} / 0.94 & \textbf{0.82} & 8.77 & 0.60 \\
mSSL-DCPE & 0.73 / \textbf{1.00} & \textbf{0.99} / 0.94 & 0.82 & 8.93 & 0.14 \\
MRCE & \textbf{0.90} / 0.66 & 0.40 / 0.70 & 0.45 & 13.08 & 592.64 \\
CAPME & 0.86 / 0.75 & 0.47 / 0.78 & 0.51 & 23.07 & 136.29 \\
SEP.L+G & 0.85 / 0.84 & 0.57 / 0.84 & 0.60 & 17.17 & 2.43 \\
SEP.SSL+SSG &  0.73 / \textbf{1.00} & \textbf{0.99} / \textbf{0.95} & \textbf{0.82} & \textbf{8.47} & \textbf{0.07} \\ \hline
\end{tabular}

\end{table}

\begin{table}[H]
\centering
\singlespacing
\caption{Average variable selection performance in the high-dimensional setting. MSE has been re-scaled by a factor of 1000. NaN indicates that the specified quantity was undefined, either because no non-zero estimates were returned or because there were truly no non-zero parameters (Simulation 8).}
\label{tab:hi_dim_var}
\begin{tabular}{lccccc}\hline
Method & SEN/SPE & PREC/ACC & MCC & FROB & TIME \\ \hline
\multicolumn{6}{c}{Simulation 5: $n = 400, p = 500, q = 25, \rho = 0.9$} \\ \hline
mSSL-DPE & \textbf{0.95} / \textbf{1.00} & \textbf{1.00} / \textbf{0.99} & \textbf{0.96} & \textbf{0.41} & 2229.21 \\
mSSL-DCPE & 0.88 / \textbf{1.00} & 0.99 / 0.97 & 0.92 & 1.40 & 23.66 \\
MRCE & 0.40 / 0.63 & 0.67 / 0.59 & 0.07 & 171.73 & 7116.94 \\
CAPME & \textbf{0.95} / 0.54 & 0.34 / 0.62 & 0.40 & 8.49 & 7625.05 \\
SEP.L+G& 0.92 / 0.76 & 0.49 / 0.79 & 0.56 & 10.33 & 19.21 \\
SEP.SSL+SSG& 0.88 / \textbf{1.00} & 0.98 / 0.97 & 0.91 & 2.25 & \textbf{3.14} \\ \hline
\multicolumn{6}{c}{Simulation 6: $n = 400, p = 500, q = 25, \rho = 0.7$} \\ \hline
mSSL-DPE & 0.91 / \textbf{1.00} & \textbf{0.99} / \textbf{0.98} & \textbf{0.94} & \textbf{1.19} & 2260.99 \\
mSSL-DCPE & 0.88 / \textbf{1.00} & \textbf{0.99} / 0.97 & 0.92 & 1.65 & 23.93 \\
MRCE & 0.74 / 0.30 & 0.33 / 0.39 & 0.07  & 87.43 & 9092.95 \\
CAPME & 0.68 / 0.84 & 0.53 / 0.81 & 0.48 & 109.03 & 7243.94 \\
SEP.L+G & \textbf{0.92} / 0.76 & 0.48 / 0.79 & 0.56 & 10.26 & 19.11 \\
SEP.SSL+SSG & 0.88 / \textbf{1.00} & 0.98 / 0.97 & 0.91 & 2.22 & \textbf{3.11} \\ \hline
\multicolumn{6}{c}{Simulation 7: $n = 400, p = 500, q = 25, \rho = 0.5$} \\ \hline
mSSL-DPE & 0.91 / 0.61 & 0.39 / 0.67 & 0.43 & 33.52 & 3839.74 \\
mSSL-DCPE & 0.88 / \textbf{1.00} & \textbf{0.99} / \textbf{0.97} & \textbf{0.92} & \textbf{1.92} & 24.12 \\
MRCE & 0.65 / 0.36 & 0.39 / 0.42 & 0.04 & 107.27 & 9540.04 \\
CAPME & 0.66 / 0.86 & 0.54 / 0.82 & 0.48 & 116.39 & 7594.80 \\
SEP.L+G & \textbf{0.92} / 0.76 & 0.49 / 0.79 & 0.56 & 10.23 & 19.28 \\
SEP.SSL+SSG & 0.88 / \textbf{1.00} & 0.98 / \textbf{0.97} & 0.91 & 2.22  & \textbf{3.14} \\ \hline
\multicolumn{6}{c}{Simulation 8: $n = 400, p = 500, q = 25, \rho = 0$} \\ \hline
mSSL-DPE & 0.91 / 0.58 & 0.35 / 0.64 & 0.39 & 36.26 & 2800.63 \\ 
mSSL-DCPE & 0.88 / \textbf{1.00} & \textbf{0.98} / \textbf{0.97} & \textbf{0.91} & 2.25 & 23.82 \\
MRCE & 0.59 / 0.41 & 0.42 / 0.45 & 0.03 & 123.23 & 9187.28 \\
CAPME & 0.66 / 0.86 & 0.54 / 0.82 & 0.48 & 116.36 & 7255.42 \\
SEP.L+G & \textbf{0.92} / 0.76 & 0.49 / 0.79 & 0.56 & 10.27 & 19.26 \\
SEP.SSL+SSG & 0.88 / \textbf{1.00} & \textbf{0.98} / \textbf{0.97} & \textbf{0.91} & \textbf{2.24} & \textbf{3.22} \\ \hline
\end{tabular}
\end{table}

\begin{table}[H]
\centering
\singlespacing
\caption{Average covariance selection performance in the low-dimensional setting. NaN indicates that the specified quantity was undefined, either because no non-zero estimates were returned or because there were truly no non-zero parameters (Simulation 4).}
\label{tab:hi_dim_cov}
\begin{tabular}{lccccc}\hline
Method & SEN/SPE & PREC/ACC & MCC & FROB & TIME \\ \hline
\multicolumn{6}{c}{Simulation 1: $n = 100, p = 50, q = 25, \rho = 0.9$} \\ \hline
mSSL-DPE & 0.97 / \textbf{0.99} & \textbf{0.92} / \textbf{0.99} & \textbf{0.94} & \textbf{167.29} & 10.23 \\
mSSL-DCPE & 0.79 / 0.96 & 0.62 / 0.94 & 0.67 & 1130.89 & 0.37 \\
MRCE & 0.96 / 0.73 &  0.24 / 0.75 & 0.41 & 675.19 & 1467.64 \\
CAPME & \textbf{1.00} / 0.00 & 0.08 / 0.08 & NaN & 2292.72 & 133.26 \\
SEP.L+G  & \textbf{0.99} / 0.67 & 0.21 / 0.69 & 0.37 & 2502.15 & 2.62 \\
SEP.SSL+SSG & 0.79 / 0.96 & 0.64 / 0.95 & 0.68 & 1456.17 & \textbf{0.09} \\ \hline
\multicolumn{6}{c}{Simulation 2: $n = 100, p = 50, q = 25, \rho = 0.7$} \\ \hline
mSSL-DPE & \textbf{1.00} / \textbf{1.00} & \textbf{1.00} / \textbf{1.00} & \textbf{1.00} & \textbf{8.94} & 1.62 \\
mSSL-DCPE & 0.95 / \textbf{1.00} & 0.95 / 0.99 & 0.94 & 28.44 & 0.21 \\
MRCE & \textbf{1.00} / 0.80 & 0.34 / 0.82 & 0.52 & 21.76 & 1704.69 \\
CAPME & 0.96 / 0.43 & 0.13 / 0.47 & 0.23  & 90.64 & 137.88 \\
SEP.L+G & \textbf{1.00} / 0.78 & 0.29 / 0.80 & 0.47 & 139.80 & 2.60 \\
SEP.SSL+SSG & 0.94 / \textbf{1.00} & 0.95 / 0.99 & 0.94 & 39.76 & \textbf{0.07} \\ \hline
\multicolumn{6}{c}{Simulation 3: $n = 100, p = 50, q = 25, \rho = 0.5$} \\ \hline
mSSL-DPE & 0.89 / \textbf{1.00} & 0.98 / \textbf{0.99} & \textbf{0.93} & \textbf{6.13} & 1.28 \\
mSSL-DCPE & 0.28 / \textbf{1.00} & 0.99 / 0.94 & 0.73 & 22.90 & 0.16 \\
MRCE & \textbf{1.00} / 0.82 & 0.32 / 0.83 & 0.51 & 7.22 & 714.19 \\
CAPME & 0.98 / 0.38 & 0.12 / 0.43 & 0.21 & 15.79 & 138.30 \\
SEP.L+G & 0.97 / 0.83 & 0.34 / 0.84 & 0.52 & 25.37 & 2.56 \\
SEP.SSL+SSG & 0.57 / \textbf{1.00} & \textbf{0.99} / 0.97 & 0.74 & 13.91 & \textbf{0.06} \\ \hline
\multicolumn{6}{c}{Simulation 4: $n = 100, p = 50, q = 25, \rho = 0$} \\ \hline
mSSL-DPE& NaN / \textbf{1.00}  & NaN / \textbf{1.00} & NaN & \textbf{0.92} & 0.60 \\
mSSL-DCPE & NaN / \textbf{1.00} & NaN / \textbf{1.00} & NaN & 0.70 & 0.14 \\
MRCE & NaN  / 0.97 &  0.00 / 0.97 & NaN & 6.27 & 592.64 \\
CAPME & NaN / 0.43 & 0.00 /  0.43 & NaN & 7.55& 136.29 \\
SEP.L+G & NaN / 0.85 & 0.00 / 0.85 & NaN & 1.24 & 2.43 \\
SEP.SSL+SSG& NaN / \textbf{1.00} & NaN / \textbf{1.00} & NaN & 0.70 & \textbf{0.07} \\ \hline
\end{tabular}
\end{table}

\begin{table}[H]
\centering
\singlespacing
\caption{Average covariance selection performance in the high-dimensional setting. NaN indicates that the specified quantity was undefined, either because no non-zero estimates were returned or because there were truly no non-zero parameters (Simulation 8).}
\label{tab:hi_dim_cov}
\begin{tabular}{lccccc}\hline
Method & SEN/SPE & PREC/ACC & MCC & FROB & TIME \\ \hline
\multicolumn{6}{c}{Simulation 5: $n = 400, p = 500, q = 25, \rho = 0.9$} \\ \hline
mSSL-DPE & 0.97 / 0.98 & \textbf{0.84} / \textbf{0.98} & \textbf{0.89} & \textbf{97.92} & 2229.21 \\
mSSL-DCPE & \textbf{1.00} / 0.89 & 0.45 / 0.90 & 0.63 & 1226.78  & 23.66 \\
MRCE & 0.94 / 0.22 & 0.11 / 0.28 & 0.21 & $6.17 \times 10^{6}$ & 7116.94 \\ 
CAPME & 0.00 / \textbf{1.00} & NaN / 0.92 & NaN & 2989.33 & 7625.05 \\
SEP.L+G & \textbf{1.00} / 0.60 & 0.18 / 0.63 & 0.33 & 2682.86 & 19.21 \\
SEP.SSL+SSG & 0.99 / 0.87 & 0.41 / 0.88 & 0.59 & 1953.69 & \textbf{3.14} \\ \hline
\multicolumn{6}{c}{Simulation 6: $n = 400, p = 500, q = 25, \rho = 0.7$} \\ \hline
mSSL-DPE & 0.99 / \textbf{1.00} & \textbf{0.95} / \textbf{1.00} & \textbf{0.97} & 22.10 & 2260.99 \\
mSSL-DCPE & \textbf{1.00} / 0.96 & 0.72 / 0.97 & 0.83 & \textbf{14.36} & 23.93 \\
MRCE & 0.92 / 0.45 & 0.16 / 0.49 & 0.28 & $16.16 \times 10^{6}$  & 9092.95 \\ 
CAPME & 0.00 / \textbf{1.00}  & NaN / 0.92 & NaN & 285.86 & 7243.94 \\
SEP.L+G & 0.99 / 0.87 & 0.40 / 0.88 & 0.58 & 161.84 & 19.11 \\
SEP.SSL+SSG & \textbf{1.00} / 0.96 & 0.71 / 0.97 & 0.83 & 57.68 & \textbf{3.11} \\ \hline
\multicolumn{6}{c}{Simulation 7: $n = 400, p = 500, q = 25, \rho = 0.5$} \\ \hline
mSSL-DPE & 0.07 / \textbf{1.00} & 0.95 / 0.93 & 0.50 & $3.59 \times 10^{4}$ & 3839.74 \\
mSSL-DCPE & \textbf{1.00} / \textbf{1.00} & 0.97 / \textbf{1.00} & 0.98 & \textbf{2.18} & 24.12 \\
MRCE & 0.87 / 0.49 & 0.17 / 0.52 & 0.32 & $1.15 \times 10^{9}$ & 9540.04 \\
CAPME & 0.00 / 1.00 & NaN / 0.92 & NaN & 87.10 & 7594.80 \\
SEP.L+G  & 0.86 / 0.96 & 0.66 / 0.95 & 0.72 & 29.30 & 19.28 \\ 
SEP.SSL+SSG & \textbf{1.00} / \textbf{1.00} & \textbf{0.98} / \textbf{1.00} & \textbf{0.99} & 4.38 & \textbf{3.14} \\ \hline
\multicolumn{6}{c}{Simulation 8: $n = 400, p = 500, q = 25, \rho = 0$} \\ \hline
mSSL-DPE & NaN / \textbf{1.00} & NaN / \textbf{1.00} & NaN & $4.03 \times 10^{4}$ & 2800.63 \\ 
mSSL-DCPE & NaN / \textbf{1.00} & NaN / \textbf{1.00} & NaN & 1.14 & 23.82 \\
MRCE & NaN / 0.46 & 0.00 / 0.46 & NaN & $5.07 \times 10^{9}$ & 9187.28 \\
CAPME & NaN / \textbf{1.00} & NaN / \textbf{1.00} & NaN & 24.00 & 7255.42 \\
SEP.L+G & NaN / 0.98 & 0.00 / 0.98 & NaN & \textbf{0.49} & 19.26 \\
SEP.SSL+SSG & NaN / \textbf{1.00} & 0.00 / \textbf{1.00} & NaN & 1.13 & \textbf{3.22} \\ \hline
\end{tabular}
\end{table}

In both the high- and low-dimensional settings, we see immediately that the regularization methods utilizing cross-validation (MRCE, CAPME, and SEP.L+G) are characterized by high sensitivity, moderate specificity, and low precision in recovering the support of both $B$ and $\Omega.$
The fact that the precisions of these three methods are less than 0.5 highlights the fact that the majority of the non-zero estimates returned are in fact false positives, a rather unattractive feature from a practitioner's standpoint! 
This is not entirely surprising, as cross-validation has a well-known tendency to over-select. 
In stark contrast are mSSL-DPE, mSSL-DCPE, and SEP.SSL+SSG, which all utilized adaptive spike-and-slab penalties.
These methods are all characterized by somewhat lower sensitivity than their cross-validated counterparts but with vastly improved specificity and precision, performing exactly as anticipated by \citet{RockovaGeorge2016}'s simulations from the univariate setting. 
In a certain sense, the regularization competitors cast a very wide net in order to capture most of the non-zero parameters, while our methods are much more discerning. 
So while the latter methods may not capture as much of the true signal as the former, they do not admit nearly as many false positives.

CAPME, SEP.L+G, and SEP.SSL+SSG all estimate $B$ in a column-wise fashion and are incapable of ``borrowing strength'' across outcomes.
MRCE and mSSL-DPE are the only two methods considered that explicitly leverage the residual correlation between outcomes from the outset. 
As noted above, in the low-dimensional settings, MRCE tended to over-select in $B$ and $\Omega,$ leading to rather poor estimates of both matrices.
In all but Simulations 7 and 8, mSSL-DPE displayed far superior estimation and support recovery performance than MRCE.

Recall that mSSL-DCPE proceeds by finding a conditional mode $(\hat{B}^{L}, \hat{\theta}^{L})$ fixing $\Omega = I,$ finding a conditional mode $(\hat{\Omega}^{L}, \hat{\eta}^{L})$ fixing $B = \hat{B}^{L},$ and then refining these two conditional modes to a single joint mode. 
It is only in this last refining step that mSSL-DCPE considers the correlation between residuals while estimating $B$.
As it turns out, this final refinement did little to change the estimated support of $B,$ so the nearly identical performance of SEP.SSL+SSG and mSSL-DCPE is not that surprising.
Further, the only practical difference between the two procedures is the adaptivity of the penalties on $\beta_{j,k}$: in SEP.SSL+SSG, the penalties separately adapt to the sparsity within each column of $B$ while in mSSL-DCPE, they adapt to the overall sparsity of $B.$

\begin{figure}[H]
\centering
\includegraphics[width =\textwidth]{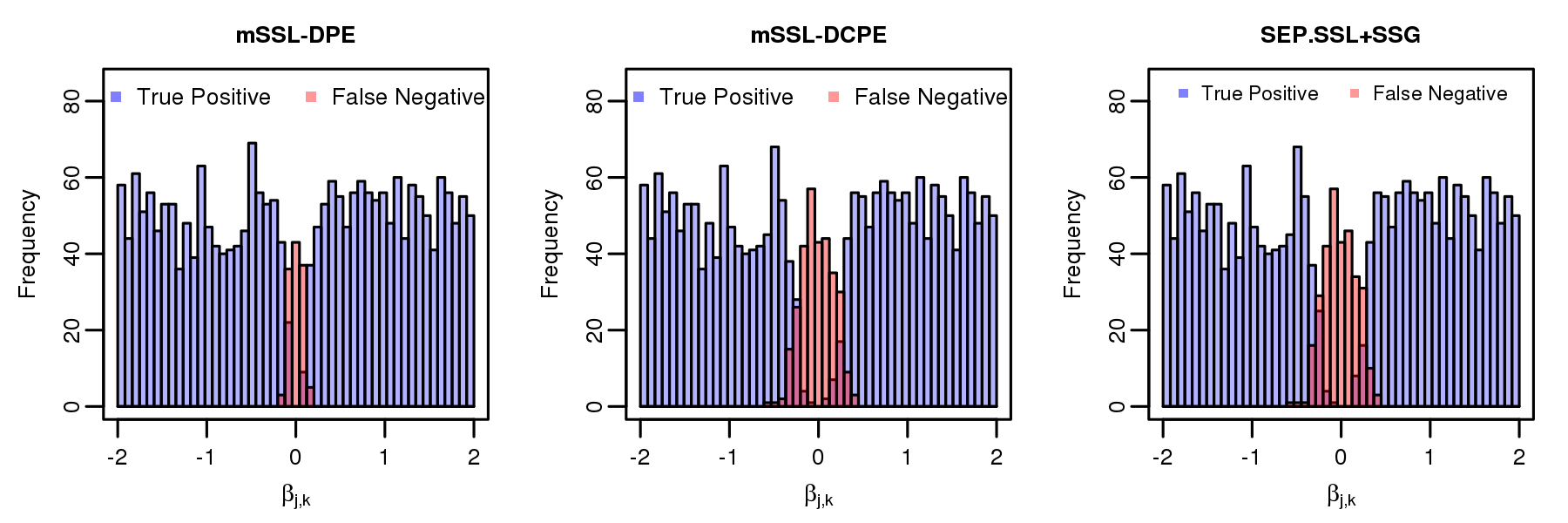}
\caption{Histograms of non-zero $\beta_{j,k}$ values that are correctly identified as non-zero (blue) and non-zero $\beta_{j,k}$ values incorrectly identified as zero (red). mSSL-DPE demonstrates the greatest acuity in recovering small $\beta_{j,k}$ values.}
\label{fig:B_histogram}
\end{figure}

By simulating the non-zero $\beta_{j,k}$'s uniformly from $\left[-2,2\right],$ we were able to compare our methods' abilities to detect signals of varying strength.
Figure~\ref{fig:B_histogram} super-imposes the distribution non-zero $\beta_{j,k}$'s correctly identified as non-zero with the distribution of non-zero $\beta_{j,k}$'s incorrectly estimated as zero by each of mSSL-DPE, mSSL-DCPE, and SEP.SSL+SSG from a single replication of Simulation 5.

In this situation, mSSL-DPE displays greater acuity for detecting smaller $\beta_{j,k}$'s than mSSL-DCPE or SEP.SSL+SSG, which are virtually ignorant of the covariance structure of the outcomes.
This is very reminiscent of \citet{Zellner1962}'s observation that multivariate estimation of $B$ in seemingly unrelated regressions is asymptotically more efficient than proceeding response-by-response and ignoring the correlation between responses.
To get a better sense as to why this may the case, recall the refined thresholding used to update our estimates of $\beta_{j,k}$ in our ECM algorithm:
$$
\beta_{j,k}^{new} = \frac{1}{n}\left(\left\lvert z_{j,k} \right\rvert - \lambda^{\star}(\beta_{j,k}^{old},\theta)\right)_{+}\text{sign}(z_{j,k})\mathbb{I}\left(\left\lvert z_{j,k} \right\rvert > \Delta_{j,k}\right).
$$
The quantity $z_{j,k}$ can be decomposed as
$$
z_{j,k} = n\beta^{old}_{j,k} + \bx_{j}'\mathbf{r}_{k}(B^{old}) + \sum_{k' \neq k}{\frac{\omega_{k,k'}}{\omega_{k,k}}\bx_{j}'\mathbf{r}_{k'}(B^{old})}.
$$
Writing $z_{j,k}$ in this way, we can readily see how $\omega_{k,k'}$ regulates the degree to which our estimate of $\beta_{j,k}$ depends on the outcome $\by_{k'}$: if $\omega_{k,k'}$ is close in value to $\omega_{k,k}$, our estimate of variable $j$'s impact on outcome $k$ will depend almost much as on the residuals $\mathbf{r}_{k'}$ as they do on the residuals $\mathbf{r}_{k}.$
On the other hand, if $\omega_{k,k'} = 0,$ then we are unable to ``borrow strength'' and use information contained in $\by_{k'}$ to help estimate $\beta_{j,k}.$
Non-zero values of $\omega_{k,k'}$ in the sum in the above expression may make it easier for some $z_{j,k}$'s corresponding to small $\beta_{j,k}$ values to overcome the thresholds $\Delta^{U}_{j,k}$ and $\lambda^{\star}_{j,k}$ in mSSL-DPE, resulting in far fewer false negative identifications in the support of $B$ than mSSL-DCPE.


Finally we must address Simulations 7 and 8, in which mSSL-DPE appears to perform exceptionally poorly.
On closer inspection, in all of the replications, mSSL-DPE stabilized immediately at a rather dense estimate of $B$ that left very little residual variance and produced a diagonal estimate of $\Omega$ with massive entries on the diagonal. 
As it turns out, the log-posterior evaluated at this estimate with $(\lambda_{0}, \xi_{0})  = (\lambda_{0}^{(L)}, \xi_{0}^{(L)})$ was considerably smaller than the log-posterior evaluated at mSSL-DCPE's estimate.
In other words, mSSL-DCPE was able to escape the ``dense B -- unstable, diagonal $\Omega$'' region of the parameter space and navigate to regions of higher posterior density.
In Simulation 7, the truly non-zero $\omega_{k,k'}$'s were rather small and in Simulation 8, $\Omega$ was the identity.
Taken together, these two simulations suggests that when $p > n,$ estimating $B$ and $\Omega$ jointly with small values of $\lambda_{0}$ can lead to sub-optimal estimates. 
In practice, we recommend running both mSSL-DPE and mSSL-DCPE and reporting results of whichever estimate has higher log-posterior. 

\section{Full Multivariate Analysis of the Football Safety Data}
\label{sec:real_data}

More than 1 million high school students played American-style tackle football in 2014, but many medical professionals have recently begun questioning the safety of the sport \citep{Bachynski2016, Pfister2016} or called for its outright ban \citep{MilesPrasad2016}.
Concern over the long-term safety of the sport have been driven partially by studies like \citet{Lehman2012}, which found an increased risk of neurodegenerative disease and \citet{Guskiewicz2005, Guskiewicz2007} and \citet{Hart2013}, which highlighted associations between concussion history and later-life cognitive impairment and depression. 

In a recent observational study, \citet{DeshpandeHasegawa2017} studied the effect of playing high school football on later-life cognitive and mental health using data from the Wisconsin Longitudinal Study (WLS), which has followed 10,317 people since they graduated from a Wisconsin high school in 1957.
In addition to an indicator of participation in high school football, the WLS dataset contains a rich set of baseline variables that may be associated with later-life health, including adolescent IQ, percentile rank in high school, and anticipated years of education.
Further, the WLS dataset contains many socio-economic outcomes measured in the mid-1970's, when the participants were in their mid-to-late 30's, as well as results from a battery of cognitive, psychological, and behavioral tests conduced in 1993, 2003-05, and 2011, when the subjects were approximately 54, 65, and 72 years of age.
\citet{DeshpandeHasegawa2017} took a univariate approach, analyzing each outcome separately, and found no evidence of a harmful effect of playing high school football on any outcome considered, after carefully adjusting for several important confounders.

We now re-visit the dataset of \citet{DeshpandeHasegawa2017} from a full multivariate perspective with mSSL-DPE and mSSL-DCPE.
Our more powerful multivariate methodology not only confirms the main findings of their analysis but also provides new insight into the residual inter-dependence of the cognitive, psychological, and socio-economic outcomes that was otherwise unavailable in their univariate analysis. 

In order to isolate the effect of playing football, \citet{DeshpandeHasegawa2017} began by creating matched sets containing one football player and one or more control subjects, or one control subject and one or more football players, using full matching with a propensity score caliper. 
These matched sets optimally balance the distribution of each baseline variable between football players and controls, and were constructed in such a way that the standardized difference in means between the two groups was less than 0.2 standard deviations.
They then regressed several standardized cognitive, psychological, behavioral, and socio-economic outcomes onto the indicator of football participation, the baseline covariates, and indicator variables for matched set inclusions.
This allowed them to estimate the effect of playing football with the associated partial slope. 
This combination of full matching and model-based covariate adjustment has been shown to remove biases due to residual covariate imbalance \citep{CochranRubin1973, SilberRosenbaum2001} in an efficient and robust fashion \citep[see, e.g.,][]{Rosenbaum2002, Hansen2004,Rubin1973, Rubin1979}.

The cognitive outcomes considered included scores on Letter Fluency (LF), Immediate Word Recall (IWR), Delayed Word Recall (DWR) , Digit Ordering (DO), WAIS Similarity (SIM), and Number Series (NS) tests.
All of these tests were administered in both 2003 and 2011, except for SIM which was also administered in 1993 and NS which was only administered in 2011. 
The psychological and behavioral outcomes included scores on the Center for Epidemiological Studies-Depression scale (CES-D), Anger Index (ANG), Hostility Index (HOS), and Anxiety Index (ANX).
CES-D and HOS scores were available from 1993, 2003, and 2011, while ANG and ANX scores were available only in 2003 and 2011. 
The socio-economic and education outcomes included occupational prestige scores (SEI) for jobs held in 1964, 1970, 1974, and 1975, number of weeks worked (WW) in 1974, earnings (EARN) in 1974, and number of years of education completed by 1974. 

We now focus on the $n = 448$ subjects with all available outcomes. 
Of these 448 subjects, 157 played high school football. 
Following the broad outline of \citet{DeshpandeHasegawa2017}, we first matched football players to controls along several baseline covariates using full matching and a propensity caliper.
Table~\ref{tab:wls_covariates} lists these covariates, along with their pre- and post-matching means and standardized differences for the football players and controls.
In all we had 157 matched sets, each comprised of a single football player and up to 6 controls, that adequately balanced the distribution of each baseline covariate. 
We then standardized each of the $q = 29$ outcomes and regressed them onto the $p = 204$ predictors, which included all of the covariates listed in Table~\ref{tab:wls_covariates} as well as indicators of matched set inclusion. 
Like the simulation study in Section~\ref{sec:simulations}, we ran mSSL-DPE and mSSL-DCPE with $\mathcal{I}_{\lambda}$ and $\mathcal{I}_{\xi}$ containing 10 evenly spaced points ranging from $1$ to $n$ and $0.1n$ to $n$, respectively, and set $a_{\theta} = a_{\eta} = 1, b_{\theta} = pq =  5,916$ and $b_{\eta} = q = 29.$ 

\begin{table}[H]
\centering
\tiny
\caption{Baseline covariates, along with pre- and post-matching means and standardized differences}
\label{tab:wls_covariates}
\begin{tabular}{lccccc} \hline
~ & ~ & \multicolumn{2}{c}{Control Mean} & \multicolumn{2}{c}{Standardized Differences} \\ \cline{3-6}
Covariate & FB Mean & Pre-Match& Post-Match & Pre-Match & Post-Match \\ \hline
Occupational Prestige of Job Aspired To in 1954 & 581.97 & 523.52 & 555.55 & 0.25 & 0.11 \\
High School Size & 138.08 & 179.92 & 146.24 & -0.33 & -0.06 \\
High School Rank (quantile) & 55.81 & 44.56 & 51.94 & 0.43 & 0.15 \\
Considered outstanding by teacher (\%) & 13 & 9 & 12 & 0.13 & 0.04 \\
Parental Income (\$100) & 73.19 & 59.63 & 59.49 & 0.19 & 0.19 \\
Participated in band or orchestra (\%) & 32 & 37 & 35 & -0.09 & -0.05 \\
Participated in speech or debate (\%) & 32 & 22 & 28 & 0.25 & 0.10 \\
Participated in school publications (\%) & 25 & 15 & 22 & 0.26 & 0.08 \\
Father was a farmer (\%) & 26 & 22 & 23 & 0.10 & 0.07 \\
Planned to serve in military (\%) & 25 & 30 & 27 & -0.12 & -0.06 \\
Attended Catholic high school (\%) & 4 & 8 & 5 & -0.19 & -0.03 \\
IQ & 105.11 & 100.40 & 103.03 & 0.34 & 0.15 \\
Father's Education (years) & 9.73 & 9.40 & 9.50 & 0.10 & 0.07 \\
Mother's Education (years) & 10.80 & 10.20 & 10.66 & 0.22 & 0.05 \\
Lived with both parents (\%) & 89 & 91 & 91 & -0.07 & -0.05 \\
Mother Working in 1957 (\%) & 42 & 33 & 38 & 0.19 & 0.09 \\
Teachers Encouraged College (\%) & 63 & 45 & 57 & 0.37 & 0.12 \\
Parents Encouraged College (\%) & 66 & 59 & 62 & 0.15 & 0.09 \\
Had Friend Planning to Attending College (\%) & 39 & 34 & 35 & 0.11 & 0.08 \\
Never discussed future plans with parents (\%) & 3 & 2 & 2 & 0.01 & 0.04 \\
Sometimes discussed future plans with parents (\%) & 42 & 46 & 43 & -0.08 & -0.03 \\
Often discussed future plans with parents (\%) & 56 & 52 & 55 & 0.08 & 0.02 \\
Family wealth considerably below community average (\%) & 1 & 0 & 0 & 0.16 & 0.16 \\
Family wealth somewhat below community average (\%)  & 9 & 7 & 7 & 0.08 & 0.14 \\
Family wealth considerably around community average (\%)  & 66 & 73 & 75 & -0.16 & -0.20 \\
Family wealth somewhat above community average (\%)  & 22 & 19 & 16 & 0.08 & 0.14 \\
Family wealth considerably above community average (\%)  & 2 & 1 & 1 & 0.07 & 0.04 \\
Parents cannot financially support college education (\%) & 30 & 31 & 29 & -0.03 & 0.02 \\
Parents can financially support college education with sacrifice (\%)  & 53 & 55 & 60 & -0.03 & -0.13 \\
Parents can easily financially support college education (\%) & 17 & 14 & 12 & 0.08 & 0.15 \\ \hline
\end{tabular}

\end{table}

mSSL-DCPE recovered 9 non-zero $\beta_{j,k}$'s and 41 non-zero $\omega_{k,k'}$'s.
mSSL-DPE recovered 14 non-zero $\beta_{j,k}$'s, eight of which were identified by mSSL-DCPE.
Additionally, mSSL-DPE identified 37 of the 41 non-zero entries in $\omega_{k,k'}$'s found by mSSL-DCPE along with several more.
On closer inspection, we found that mSSL-DPE's estimated mode had a slightly larger log-posterior value than mSSL-DCPE's.
In terms of estimating the effect of playing football on these outcomes, our results comport with \citet{DeshpandeHasegawa2017}'s findings from separate univariate analyses: neither mSSL-DPE nor mSSL-DCPE identified a non-zero $\beta_{j,k}$ corresponding to football participation.
Much of the signal uncovered by mSSL-DPE is quite intuitive: adolescent IQ was a relevant predictor of scores on the digits ordering task in 2003 and the WAIS similarity task in 1993, 2003, and 2011, anticipated years of post-secondary education was a strong predictor of actual years of education completed by 1974 and the occupational prestige of subjects' job in 1964, and the occupational prestige of the jobs to which subjects aspired in high school was a relevant predictor of the occupational prestige of the jobs they actually held in 1964, 1970, 1974, and 1975.
In addition, mMEVS-DPE also selected several of the indicator variables of matched set membership.
These corresponded to matched sets containing subjects with similar covariates and propensity scores who had higher than average CES-D scores in 1993 (i.e. they displayed more depressive symptoms), higher than average earnings in 1974, or higher than average scores on the Anger Index in 2004. 

\begin{figure}[H]
\centering
\includegraphics[width = 0.8\textwidth]{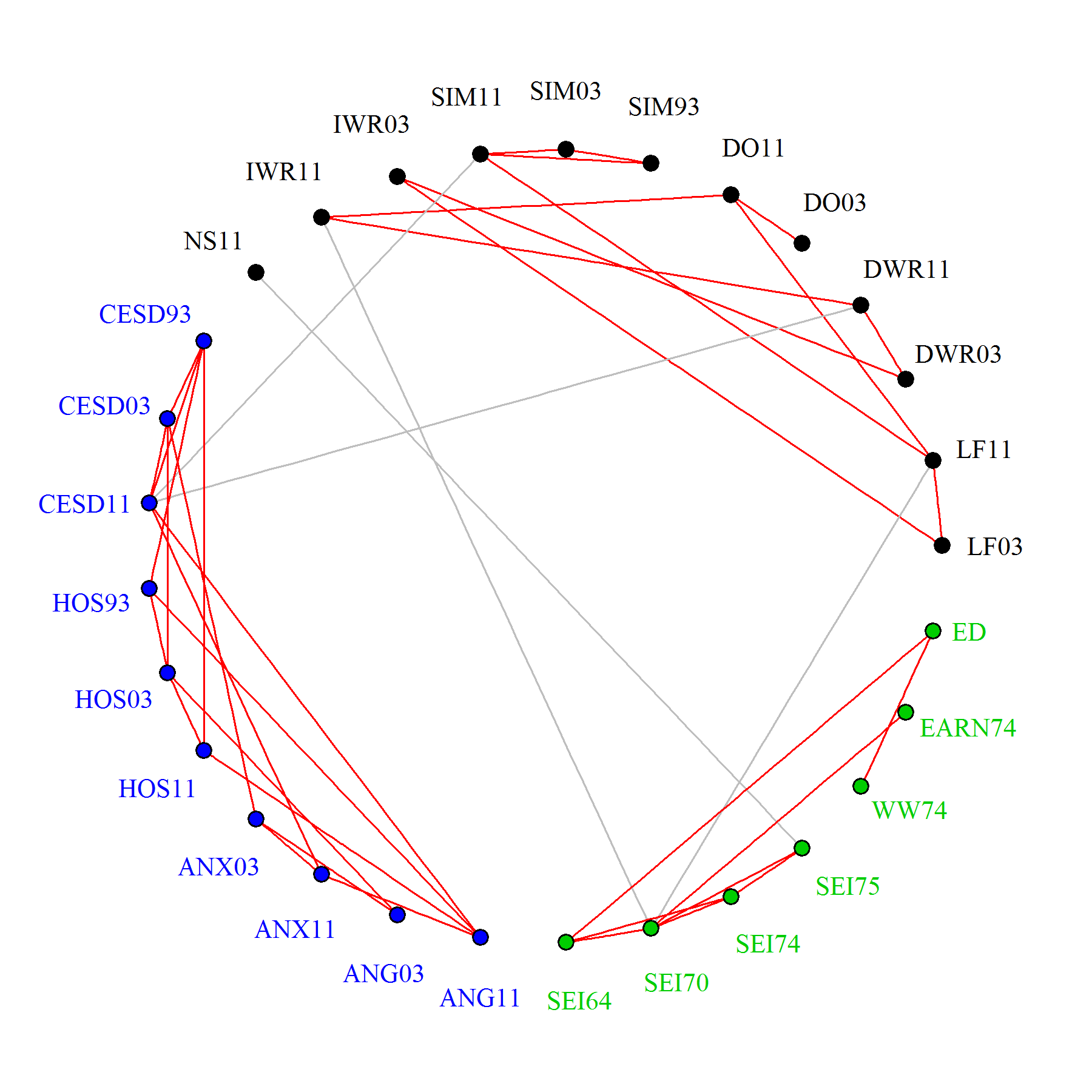}
\caption{The number following outcome abbreviation indicates the year in which it was measured. Outcomes are colored according to type: cognitive (black), psychological/behavioral (blue), socio-economic / educational (green). Observe that there are many more ``within community'' edges, colored red, than ``between community'' edges, colored gray.}
\label{fig:wls_graph}
\end{figure}

Not only does our multivariate approach confirm the main findings of \citet{DeshpandeHasegawa2017}'s univariate analysis, it also provides an estimate of the residual residual Gaussian graphical model $G$ of the 29 outcomes considered, shown in Figure~\ref{fig:wls_graph}.
The edges in $G$ encode conditional dependency between the cognitive, psychological/behavioral, and socio-economic outcomes that remain after we adjust for the measured confounders.
$G$ exhibits a very strong community structure, with many more edges between outcomes of the same type (colored in red) than of different type (colored in gray).
This is rather interesting, in light of the fact that the implicit prior on $G$, which made each edge equally likely to appear, did not tend to favor any such structure. 

Many of the conditional dependence relations represented in $G$ seem intuitive: after adjusting for the covariates listed in Table~\ref{tab:wls_covariates}, we see that results from the same cognitive test administered in multiple years tended to be conditionally dependent on each other (see, e.g., the triangle formed by SIM93, SIM03, and SIM11).
Additionally, we see that the CES-D scale depression scores and anger, hostility, and anxiety scores from the same year tended to be conditionally dependent as well.
Perhaps more interesting are the ``between community'' links between outcomes of different types, colored in red.
After adjusting for covariates, occupational prestige of the job held in 1975 (SEI75) appears conditionally dependent on the score on the number series task in 2011 (NS11), while the scores on both the CES-D scale and letter fluency test (CESD11 and LF11) are conditionally dependent on the similarity test result in 2011 (SIM11).

\section{Discussion}
\label{sec:discussion}

In this article, we have built on \citet{RockovaGeorge2014}'s and \citet{RockovaGeorge2016}'s deterministic spike-and-slab formulation of Bayesian variable selection for univariate linear regression to develop a full joint procedure for simultaneous variable and covariance selection problem in multivariate linear regression models. 
We proposed and deployed an ECM algorithm within a path-following scheme to identify the modes of several posterior distributions, corresponding to different choices of spike distributions.
This dynamic exploration of several posteriors is in marked contrast to MCMC, which attempts to characterize a single posterior. 
In our simulation experiments and analysis of the football safety data, the modal estimates identified by our dynamic posterior exploration stabilized, allowing us to report a single estimate out of the many we computed without the need for cross-validation.
Though there is no general guarantee that these trajectories will stabilize, a figure like Figure~\ref{fig:dpe_trajectories} provide a useful self-check: if one observes stabilization in the supports of $B$ and $\Omega$ and in the log-posterior, one can safely report the final mode identified.
On the other hand, if the modal estimates have not stabilized, one can simply add larger values of $\lambda_{0}$ and $\xi_{0}$ to the ladders and continue exploring.

To negotiate the dynamically changing multimodal environment, we have focused on modal estimation, at the cost of temporarily sacrificing full uncertainty quantification and posterior inference.
Assessing the variability in the estimates of mSSL-DPE remains an important problem.
One could run a general MCMC simulation starting from the final mSSL-DPE estimate.
Alternatively, the relative speed of our ECM algorithm allows it to be used within \citet{TaddyLopesGardner2016}'s recently proposed bootstrap independent Metropolis-Hasting algorithm.

As anticipated by results in \citet{RockovaGeorge2014} and \citet{RockovaGeorge2016}, our procedure tends to out-perform procedures that use cross-validation to select regularization penalties.
A key driver of the improvement is the hierarchical modeling of the uncertainty of the indicators $\bgamma$ and $\bdelta,$ which allows the penalties $\lambda^{\star}_{j,k}$ and $\xi^{\star}_{k,k'}$ in our ECM algorithm selectively shrink each $\beta_{j,k}$ and $\omega_{k,k'}.$
This is in marked contrast to regularization methods that apply the same amount of shrinkage to each $\beta_{j,k}$ and the same amount of shrinkage to each $\omega_{k,k'}.$
While we have focused on the simplest setting where the $\bgamma$'s and $\bdelta$'s are treated as exchangeable, it is straightforward to incorporate more thoughtful structured sparsity within our framework. 
For instance, if the covariates displayed a known grouping structure, we could introduce several $\theta$ parameters, one for each group, with little additional computational overhead.

\newpage
\appendix

\section{Sensitivity to hyper-parameters}
\label{sec:hyper_parameters}
To run our proposed procedures, mSSL-DPE and mSSL-DCPE, it is necessary to specify several tuning parameters: the slab penalties $\lambda_{1}$ and $\xi_{1},$ the ladders of spike penalties $\mathcal{I}_{\lambda}$ and $\mathcal{I}_{\xi},$ and the hyper-parameters for the priors on $\theta$ and $\eta,$ $a_{\theta}, b_{\theta}, a_{\eta}$ and $b_{\eta}.$
In our earlier simulation study we took $\lambda_{1} = 1, \xi_{1} = 0.01n$ and let $\mathcal{I}_{\lambda}$ and $\mathcal{I}_{\xi}$ contain $L  = 10$ equally spaced values ranging from $10$ to $n$ and $0.1n$ to $n$, respectively. 
While the overall performance of our procedures with these penalty parameters is quite promising, these choices are somewhat arbitrary.
To investigate the sensitivity of our results to these choices, we now compare several alternative specifications systematically using the same simulated datasets from Simulation 1 above.

Like in the previous section, we will keep $\lambda_{1} = 1$ and also consider ladders of size $L = 10.$
Moreover, we also set the minimal spike penalty $\lambda_{0}^{(1)}$ to be 10 just as before. 
Rather than taking $\lambda_{0}^{(L)} = n,$ we now consider for comparison setting the terminal spike penalty $\lambda_{0}^{(L)} = \max_{j,k}{\left\{\left\lvert \bx_{j}^{\top}\by_{k}\right\rvert\right\}}$ to be the maximal absolute covariance between the predictors and the response.
To motivate this choice, recall the characterization of the solution $\tilde{B}$ of Equation~\eqref{eq:B_update} given by the KKT conditions:
$$
\tilde{\beta}_{j,k} = n^{-1}\left[ \left\lvert z_{j,k} \right\rvert - \lambda^{\star}(\tilde{\beta}_{j,k},  \theta)\right]_{+}\text{sign}(z_{j,k}),
$$
where
\begin{align*}
z_{j,k} &= n\tilde{\beta}_{j,k} + \sum_{k'}{\frac{\omega_{k,k'}}{\omega_{k,k}}\bx_{j}'\mathbf{r}_{k'}(\tilde{B})} \\
\lambda^{\star}_{j,k} := \lambda^{\star}(\tilde{\beta}_{j,k}, \theta) &= \lambda_{1}p^{\star}(\tilde{\beta}_{j,k}, \theta) + \lambda_{0}(1 - p^{\star}(\tilde{\beta}_{j,k}, \theta)).
\end{align*}
Observe that in the case of extreme sparsity, with $\tilde{B}$ being the zero matrix and $\Omega$ being diagonal, the argument of the soft-thresholding operator $z_{j,k}$ is just $\bx_{j}^{\top}\by_{k}.$ 
In this case, then, taking $\lambda^{\star}_{j,k} \geq \max{\left\{\left\lvert \bx_{j}^{\top}\by_{k}\right\rvert\right\}}$ ensures that all of the $\beta_{j,k}$ values remain at zero. 
Since $\lambda^{\star}_{j,k} \leq \lambda_{0},$ this argument suggests that we need not consider $\lambda_{0}$ exceeding the maximal absolute covariance between the predictors and the responses.
We note in passing that \citet{BrehenyHuang2011} use a similar argument to select maximal penalties in the univariate regression setting.

Turning our attention to $\xi_{1}$ and $\mathcal{I}_{\xi},$ we follow the example of the glasso package in \texttt{R} and consider a range of penalties based on $n^{-1}\left\lVert \bY^{\top}\bY \right\rVert_{\infty}.$
Specifically, we take $\xi^{(L)}_{0} = n^{-1}\left\lVert \bY^{\top}\bY \right\rVert_{\infty}/10, \xi^{(1)}_{0} = \xi^{(L)}/10$ and $\xi_{1} = \xi^{(1)}_{0}/10.$
Having selected the upper and lower limits of the ladders $\mathcal{I}_{\lambda}$ and $\mathcal{I}_{\xi},$ we also consider two possible specifications of the intermediate values using equally spaced values on the absolute scale and on the log-scale.

For the Beta hyper-parameters, recall from above that we took $(a_{\theta}, b_{\theta}) = (1,pq)$ and $(a_{\eta}, b_{\eta}) = (1,q).$
In the simulations above, we observed that with this specification we achieved reasonably good support recovery of the true sparse $B$ and $\Omega.$
The extent to which our prior specification drove this recovered sparsity is not immediately clear. 
Put another way, we may wonder whether our sparse estimates of $B$ and $\Omega$ truly ``discovered'' or whether they ``manufactured'' by the prior concentrating on sparse matrices? 
To probe this question, we consider setting $(a_{\theta}, b_{\theta}) = (1,p)$ and $(1,1)$ and setting $(a_{\eta}, b_{\eta}) = (1,1).$
In all, we have twelve combinations of hyper-parameter settings, which we summarize below, and compare them using the same simulated datasets as in Simulation 1 above.
\begin{description}
\item[Setting 1]{We fix $\lambda_{1} = 1, \xi_{1} = 0.01n$ and let $\mathcal{I}_{\lambda}$ and $\mathcal{I}_{\xi}$ contain $10$ equally spaced values between 10 and $n$ and $0.1n$ to $n$, respectively. We also set $b_{\theta} = pq, b_{\eta} = q$}
\item[Setting 2]{Same as Setting 1 but with $b_{\theta} = p$}
\item[Setting 3]{Same as Setting 1 but with $b_{\theta} = 1$ and $b_{\eta} = 1.$}
\item[Setting 4]{We fix $\lambda_{1} = 1, \xi_{1} = 0.01n$ and let $\mathcal{I}_{\lambda}$ and $\mathcal{I}_{\xi}$ contain 10 values equally spaced on the log-scale between $10$ and $n$ and $0.1n$ and $n,$ respectively. We also set $b_{\theta} = pq$ and $b_{\eta} = q.$}
\item[Setting 5]{Same as Setting 4 but with $b_{\theta} = p$}
\item[Setting 6]{Same as Setting 4 but with $b_{\theta} = 1, b_{\eta} = 1.$}
\item[Setting 7]{We fix $\lambda_{1} = 1$ and set $\xi_{1} = n^{-1}\lVert \bY^{\top}\bY \rVert_{\infty}/1000$. We then let $\mathcal{I}_{\lambda}$ contain 10 equally spaced values between 10 and $\lVert \bX^{\top}\bY \rVert_{\infty}$ and let $\mathcal{I}_{\xi}$ contain 10 equally spaced values between $10\xi_{1}$ and $100\xi_{1}$. We also set $b_{\theta} = pq$ and $b_{\eta} = q.$}
\item[Setting 8]{Same as Setting 7 but with $b_{\theta} = p$}
\item[Setting 9]{Same as Setting 7 but with $b_{\theta} = b_{\eta} = 1$}
\item[Setting 10]{We fix $\lambda_{1} = 1$ and set $\xi_{1} = n^{-1}\lVert \bY^{\top}\bY \rVert_{\infty}/1000$. We then let $\mathcal{I}_{\lambda}$ contain 10 values between 10 and $\lVert \bX^{\top}\bY \rVert_{\infty}$ equally spaced on the log-scale and let $\mathcal{I}_{\xi}$ contain 10 values between $10\xi_{1}$ and $100\xi_{1}$ equally spaced on the log-scale. We also set $b_{\theta} = pq$ and $b_{\eta} = q.$}
\item[Setting 11]{Same as Setting 10 but with $b_{\theta} = p$}
\item[Setting 12]{Same as Setting 10 but with $b_{\theta} = 1, b_{\eta} = 1$}
\end{description}


Tables~\ref{tab:hp_sim1_var} and~\ref{tab:hp_sim1_cov} summarize the variable selection and covariance selection performance of mSSL-DPE with these different hyper-parameter settings, respectively.
\begin{table}[H]
\centering
\singlespacing
\caption{Sensitivity of mSSL-DPE's variable selection performance to different hyper-parameter specifications. As in Tables~\ref{tab:low_dim_var} and~\ref{tab:hi_dim_var}, MSE has been re-scaled by a factor of 1000}
\label{tab:hp_sim1_var}
\begin{tabular}{lccccc} \\\hline
~ & SEN/SPE & PREC/ACC & MCC & MSE & TIME \\ \hline
Setting 1 & 0.86 / \textbf{1.00} & \textbf{1.00} / \textbf{0.97} & 0.91 & 1.66 & 11.08 \\
Setting 2 & 0.87 / \bf{1.00} & 0.99 / \textbf{0.97} & \textbf{0.92} & 1.56 & \bf{10.63} \\
Setting 3 & 0.87 / \bf{1.00} & 0.99 /  \textbf{0.97} & \textbf{0.92} & 1.55 & 10.64 \\ \hline
Setting 4 & 0.87 / \bf{1.00} & \textbf{1.00} / \textbf{0.97} & 0.91 & \bf{1.44} & 20.09 \\
Setting 5 & \bf{0.88} / \bf{1.00} & 0.99 / \textbf{0.97} & \textbf{0.92} & 1.47 & 20.26 \\
Setting 6 & \bf{0.88} / \bf{1.00} & 0.99 / \textbf{0.97} & \textbf{0.92} & 1.47 & 19.04 \\  \hline \hline
Setting 7 & 0.68 / 0.99 & 0.97 / 0.93 & 0.77 & 30.66 & 98.24 \\
Setting 8 & 0.69 / 0.99 & 0.96 / 0.93 & 0.78 & 30.04 & 114.92 \\
Setting 9 & 0.69 / 0.99 & 0.96 / 0.93 & 0.78 & 29.96 & 126.64 \\ \hline
Setting 10 & 0.86 / \bf{1.00} & 0.99 / \textbf{0.97} & 0.91 & 3.11 & 178.66 \\
Setting 11 & 0.87 / \bf{1.00} & 0.99 / \textbf{0.97} & 0.91 & 2.92 & 155.50 \\
Setting 12 & 0.87 / \bf{1.00} & 0.99 / \textbf{0.97} & 0.91 & 2.90 & 158.53 \\ \hline
\end{tabular}
\end{table}
\begin{table}[H]
\centering
\singlespacing
\caption{Sensitivity of mSSL-DPE's covariance selection performance to different hyper-parameter specifications. As in Tables~\ref{tab:low_dim_var} and~\ref{tab:hi_dim_var}, MSE has been re-scaled by a factor of 1000}
\label{tab:hp_sim1_cov}
\begin{tabular}{lccccc} \\\hline
~ & SEN/SPE & PREC/ACC & MCC & FROB & TIME \\ \hline
Setting 1 & 0.97 / \textbf{0.99} & 0.92 / \textbf{0.99} & 0.94 & 167.29 & 11.08 \\
Setting 2 & 0.98 / \textbf{0.99} & 0.93 / \textbf{0.99} & \textbf{0.95} & 140.98 & \textbf{10.63} \\
Setting 3 & 0.98 / \textbf{0.99} & 0.93 / \textbf{0.99} & \textbf{0.95} & 142.61 & 10.64 \\ \hline
Setting 4 & 0.98 / \textbf{0.99} & 0.93 / \textbf{0.99} & \textbf{0.95} & 148.41 & 20.09 \\
Setting 5 & 0.98 / \textbf{0.99} & \textbf{0.94} / \textbf{0.99} & \textbf{0.95} & \textbf{135.32} & 20.26 \\
Setting 6 & 0.98 / \textbf{0.99} & 0.93 / \textbf{0.99} & \textbf{0.95} & 136.75 & 19.04 \\ \hline \hline
Setting 7 & 0.85 / 0.79 & 0.28 / 0.79 & 0.41 & 1454.52 & 98.24 \\
Setting 8 & 0.84 / 0.80 & 0.29 / 0.81 & 0.42 & 1392.94 & 114.92 \\
Setting 9 & 0.84 / 0.80 & 0.29 / 0.81 & 0.42 & 1387.61 & 126.64 \\ \hline
Setting 10 &  \textbf{0.99} / 0.96 & 0.73 / 0.97 & 0.83 & 207.90 & 178.66 \\
Setting 11 & \textbf{0.99} / 0.96 & 0.74 / 0.97 & 0.84 & 215.20 & 155.50 \\
Setting 12 & \textbf{0.99} / 0.97 & 0.74 / 0.97 & 0.84 & 207.32 & 158.53 \\ \hline
\end{tabular}
\end{table}

Recall that in Settings 1 -- 3 we used the same spike and slab penalties as the simulations in the previous subsections, but varied the hyper-parameters $b_{\theta}$ and $b_{\eta}$ controlling the prior degree of sparsity in $B$ and $\Omega,$ respectively.
We see that the variable selection performance is very comparable between Settings 1, 2, and 3, with only slight improvements in SEN and MCC. 
Our estimation of $B$ seems to improve when we decrease $b_{\theta}$ from $pq$ to $p$ to 1.
In Setting 1, the choice $b_{\theta} = pq$ centers most of its prior probability on matrices $B$ with only a handful of large entries, while the choice $b_{\theta} = p$ places more prior probability on $B$'s with a handful of large entries in each column.
In Setting 3, the choice $b_{\theta} = 1$ centers the prior around matrices in which half of the entires are large.
Because Settings 2 and 3 center the prior closer to the underlying $B$, which contained 20\% non-zero entries, this improvement in estimation is not entirely surprising.
Comparing the results from Settings 1 -- 3 to the results from Settings 4 -- 6, we see that using spike penalties equally spaced on the log-scale yielded even better estimates of $B$ than using spike penalties equally spaced on the absolute scale.
We should point out, however, that this improvement is marginal; as the MSE reported in Table~\ref{tab:hp_sim1_var} has been rescaled by a factor of 1000, the improvement in estimation error between Settings 1 and 4 is only the order of $10^{-4}.$ 
Looking at Table~\ref{tab:hp_sim1_cov}, we see that the improved estimation of $B$ was accompanied by improved estimation of $\Omega.$

To assess the sensitivity of our results to the choice of spike and slab penalties, we may compare Settings 1 -- 3 to Settings 7 -- 9 and Settings 4 -- 6 to Settings 10 -- 12.
We immediately find that taking $\lambda_{0}^{(L)} = \max_{j,k}\left\lvert \bx_{j}^{\top}\by_{k}\right\rvert, \xi_{0}^{(L)} = n^{-1}\left\lVert \bY^{\top}\bY\right\rVert_{\infty}/10, \xi_{0}^{(1)} = \xi_{0}^{(L)}/10$ and $\xi_{1} = \xi_{0}^{(1)}/10$ yields less sensitivity support recovery of $B$ and less precise support recovery of $\Omega.$ 
In other words, with these alternative specifications, we tended to identify fewer of the non-zero $\beta_{j,k}$'s and identified many false positives in the support of $\Omega.$ 
This is accompanied by substantially worse estimation error. 
Comparing Settings 10 -- 12 to Settings 7 -- 9, it does appear that the worse support recovery is mitigated slightly by evenly spacing the spike penalties on the log-scale rather than the absolute scale.

Repeating this comparison in the remaining seven simulation settings yielded similar results: our procedures are somewhat more sensitive to the choice of the spike and slab penalties than they are to the hyper-parameters $b_{\theta}$ and $b_{\eta}.$
Based on this, we recommend setting $(\lambda_{1}, \lambda_{0}^{(1)}, \lambda_{0}^{(L)}) = (1, 10, n)$ and $(\xi_{1}, \xi_{0}^{(1)}, \xi_{0}^{(L)}) = (0.01n, 0.1n, n).$
It is somewhat trickier to definitively recommend a default choice of $b_{\theta}$ and $b_{\eta},$ however.
In our experiments, we found that setting $(b_{\theta}, b_{\eta}) = (p,q)$ or $(1,1)$ resulted in improved estimation and support recovery of $B$.
This, however, was largely an expected by-product of the fact that these choices tended to center the prior on matrices with a similar sparsity pattern as $B$ than our original choice $(b_{\theta}, b_{\eta}) = (pq, q).$
In light of this, when one has real knowledge about the overall level of sparsity of $B$ and $\Omega,$ we recommend setting $b_{\theta}$ and $b_{\eta}$ in such a way that the prior means of $\theta$ and $\eta$ are consistent with this knowledge. 
In the absence of such information, however, our original specification $b_{\theta} = pq$ and $b_{\eta} = q$ should we fine. 
However, the relative speed our procedure enables one to investigate the sensitivity to such choices quickly.

\bibliographystyle{apalike}
\newpage
\bibliography{mSSL}
\end{document}